\documentclass[usenatbib]{mn2e}
\usepackage{amsmath,amssymb,amsfonts} 
\usepackage{graphics}                 
\usepackage{color}                    
\usepackage{graphicx}
\usepackage{mathrsfs}   
\usepackage{natbib}
\usepackage{hyperref}                 
\DeclareFontFamily{U}{euc}{}
\DeclareFontShape{U}{euc}{m}{n}{<-6>eurm5<6-8>eurm7<8->eurm10}{}%
\DeclareSymbolFont{AMSc}{U}{euc}{m}{n} 
\DeclareMathSymbol{\upmu}{\mathord}{AMSc}{"16} 

\newcommand{\epi}{E_{p,int}}
\newcommand{\epo}{E_{p,obs}}
\newcommand{\eiso}{E_{iso}}
\newcommand{\eg}{E_{\gamma}}
\newcommand{\liso}{L_{iso}}
\newcommand{\sbol}{S_{bol}}
\newcommand{\pbol}{P_{bol}}
\newcommand{\pf}{P_{1024}}
\newcommand{\tn}{T_{90}}
\newcommand{\tf}{T_{50}}
\newcommand{\hr}{\text{\it{HR}}_{H}}
\newcommand{\sbhr}{$\sbol-\hr$}
\newcommand{\sbep}{$\sbol-\epo$}
\newcommand{\eiep}{$\eiso-\epi$}
\newcommand{\egep}{$\eg-\epi$}
\newcommand{\pbep}{$\pbol-\epo$}

\title{The Possible Impact of GRB Detector Thresholds on Cosmological Standard Candles}
\author[A. Shahmoradi and R. J. Nemiroff]{A. Shahmoradi$^{1}$\thanks{E-mail:
ashahmor@mtu.edu; nemiroff@mtu.edu} and R. J. Nemiroff$^{1}$\footnotemark[1]\\
$^{1}$Department of Physics, Michigan Technological University, Houghton, MI 49931}

\begin{document}

\date{\date{Accepted ... Received ... ; in original form ...}}

\pagerange{\pageref{firstpage}--\pageref{lastpage}} \pubyear{2009}

\maketitle

\label{firstpage}

\begin{abstract}
GRB satellites are relatively inefficient detectors of dim hard bursts because they trigger on photon counts, which are number-biased against hard photons.  Therefore, for example, given two bursts of identical peak luminosity near the detection threshold, a dim soft burst will be preferentially detected over a dim hard burst.  This detector bias can create or skew an apparent correlation where increasingly hard GRBs appear increasingly bright.  Although such correlations may be obfuscated by a middle step where GRBs need to be bright enough to have their actual redshifts determined, it is found that the bias is generally pervasive.  This result is derived here through simulations convolving a wide variety of possible GRB brightnesses and spectra with the BATSE Large Area Detectors (LAD) detection thresholds.  The presented analyses indicate that the rest-frame $\nu F_{\nu}$ spectrum peak energy of long-duration GRBs, $\epi$, is not a good cosmological standard candle without significant corrections for selection effects.  Therefore, the appearance of $\epi$ in seeming correlations such as the Amati ($E_{iso}-\epi$), Ghirlanda ($E_{\gamma}-\epi$), and $L_{iso}-\epi$ relations is statistically real but strongly influenced by so far uncalibrated GRB detector thresholds.
\end{abstract}


\section{Introduction}
\label{sec:introduction}
The existence of correlations among the spectral parameters of Long-duration Gamma-Ray Bursts (LGRBs) has been touted as providing clues to the underling physics of GRB prompt emission and making LGRBs a useful tool for probing cosmology in the distant universe. Recently reported attempts to use these correlations to construct a GRB Hubble diagram include those by Cardone et al. (2009), Amati et al. (2008a), Basilakos \& Perivolaropoulos (2008), Cuesta et al. (2008), Liang et al. (2008), Schaefer (2007) -- hereafter S07 -- \& Firmani et al. (2006).  The investigation of possible correlations among the parameters of LGRBs, however, dates back to the BATSE era when the cosmological origins of LGRBs was not yet established. Specifically in an early effort, Lloyd, Petrosian, \& Mallozzi (2000) did an analysis to determine the degree of correlation between the observer-frame $\nu F_{\nu}$ spectrum peak energy ($\epo$) and the bolometric fluence of bright BATSE LGRBs ($\sbol$) and to investigate to what extent the correlation is either intrinsic or cosmological.  After accounting for the data truncation due to the detection threshold, they concluded that there is probably a significant correlation between the rest-frame $\nu F_{\nu}$ spectrum peak energy ($\epi$) and isotropic-equivalent radiated energy ($\eiso$).

While there is still no unique and robust interpretation of these results (e.g. Levinson \& Eichler 2005; Rees \& M\'{e}sz\'{a}ros 2005; Eichler \& Levinson 2004), the discovery of some outliers to these relations (e.g. Urata et al. 2009; Sugita et al. 2009; Bellm et al. 2008; McBreen et al. 2008; Campana  et al. 2007; Rizzuto et al. 2007; Gehrels et al. 2006; Vaghuan et al. 2006; Sazonov et al. 2004; Soderberg et al. 2004; Ghirlanda et al. 2004a -- hereafter G04a) has raised two possibilities: that these correlations belong to only a sub-population of LGRBs, or that they are an artifact of the GRB detection process. The latter idea is bolstered by recent reports from several independent groups (Butler et al. 2009a; Bagoly et al. 2009; Bagoly et al. 2008; Nava et al. 2008, hereafter N08; Butler et al. 2007, hereafter B07). Moreover, Nakar \& Piran (2005a) (hereafter NP05a), found that a significant number of BATSE LGRBs are inconsistent with two of these relations known as the `Amati relation' (Amati 2006; Amati 2002) and the `Ghirlanda relation' (Ghirlanda et al. 2007 -- hereafter G07; G04a) which relate $\epi$ of LGRBs to their isotropic-equivalent radiated energy ($\eiso$) and the collimation-corrected energy ($E_{\gamma}$) respectively.  Following their analysis, Band \& Preece (2005), analyzed a large sample of BATSE LGRBs and found that about $88\%$ and $1.6\%$ of their sample were inconsistent with the Amati and Ghirlanda relations respectively. Kaneko et al. (2006) -- hereafter K06 -- also reported an inconsistency of bright BATSE bursts with these correlations. 

Responding to these results, Ghirlanda et al. (2005a) argued that taking into account the intrinsic scatter of the Amati relation, the BATSE bursts may still be consistent.  This claim, however, has also been challenged (Nakar \& Piran 2005b). Although all the above mentioned reports generally conclude that the Amati and Ghirlanda relations are statistically non-compelling, the matter still remains controversial whether the reported correlations are completely due to selection effects in the detection process, or whether there is some real, statistically strong correlation between the LGRBs spectral parameters. It is noteworthy that Yonetoku et al. (2004) -- hereafter Y04 -- have also tried to estimate the redshifts of 745 BATSE LGRBs using a relation between $L_{iso}$ and $E_{p, int}$. However, these estimates resulted in 21 GRBs being located beyond $z > 12$, and 35 even having no solution satisfying the $L_{iso}-\epi$ relation, indicating that these 35 bursts show a large observer-frame peak energy ($E_{p,obs}$) while being very dim. In addition, Tsutsui et al. (2008) have shown that the redshifts derived from the lag-luminosity \& $L_{iso}-\epi$ for 565 BATSE LGRBs are totally inconsistent with each other.

Most recently, N08 and Ghirlanda et al. (2008) -- hereafter G08 -- have reported the triggering threshold limits for several GRB detectors, including BATSE, SWIFT, Konus-Wind, BeppoSAX and HETE-II in the plane of peak energy vs. bolometric fluence (\sbep) and peak energy vs. bolometric 1-second peak flux (\pbep) of LGRBs. They also obtain the minimum fluence limits required for spectral analysis of a burst on these planes and conclude, contrary to the previous reports, that only $6\%$ of BATSE LGRBs are certain outliers with respect to the Amati relation at $>3\sigma$, while there are apparently no outliers to the Ghirlanda relation. They also find that the slope and the distribution of LGRBs are significantly different from the slopes of the curves of the minimum fluences and peak fluxes required for triggering and spectral analysis on these planes. Their analysis, however, is limited to about 380 bright BATSE bursts analyzed by K06 and themselves which constitute only $14\%$ of the entire number of GRBs detected by BATSE. The small sample of bursts used by N08 and G08 makes it very hard to have an accurate investigation of the trigger threshold effects on the two planes of \sbep ~and \pbep. For this reason, they focus mainly on the limits imposed by the spectral analysis of LGRBs on these planes, leaving potential effects of trigger threshold limits on the distribution of bursts untreated. 

In this paper, while focusing our attention on BATSE LAD detectors, we study the trigger threshold effects on the joint distributions of the spectral parameters of BATSE GRBs using a method that avoids the difficult direct measurement of $\epo$ by substituting a much more easily found hardness ratio (Shahmoradi \& Nemiroff 2010{\it a}).  Based on $\epo$ estimates provided in Table (4) of Shahmoradi \& Nemiroff (2010{\it a}), we are then able to include $1900$ BATSE GRBs, including $\sim500$ short duration bursts (SGRBs) ($\tn \lesssim 3 ~[sec]$) and more than 1400 LGRBs ($\tn \gtrsim 3 ~[sec]$) in our simulations of BATSE LAD detectors. In order to have an accurate investigation, we perform the analysis based on two entirely different approaches to GRB classification: 

\begin{enumerate}
	\item 
	Based on the traditional GRB classification in which we label the bursts as SGRBs or LGRBs according to the duration division determined by Kouveliotou et al. (1993), but with a higher cutoff on the duration set at $\tn=3\,[sec]$ as compared to $\tn=2\,[sec]$ used therein.  The higher cutoff is used to ensure minimal contamination of BATSE LGRBs by possible SGRBs with $\tn \sim 2\,[sec]$. 
	\\
	\item
	Based on Fuzzy Clustering algorithms in which we assign each burst a probability of being LGRB or SGRB, derived from three distinct properties of Gamma-Ray Bursts: bolometric fluence ($\sbol$), spectral peak energy ($\epo$) or equivalently, hardness ratio ($\hr$) as defined by Shahmoradi \& Nemiroff (2010{\it a}), and the duration ($\tn$). A detailed description of the methodology and algorithm used for the fuzzy classification of BATSE GRBs is given by Shahmoradi (2010).
\end{enumerate}

We are then able to proceed, in simulations and subsequent analyses, to investigate the significance of the effects of triggering threshold limits on the detection and distribution of BATSE GRBs in the planes of \sbep ~as well as \pbep ~and its possible effects on the Amati and Ghirlanda relations. Specifically, we show that the current realizations of the Amati relation as a low-dispersion ($\sigma$ $\sim$ $0.2$\,{\it dex}) Log-Log relation given by Amati (2006) \& G07, are likely skewed by a large population of dim hard LGRBs that are missed due to the impossibility of a spectral analysis that determines $\epo$. The significance of the detector threshold effects on these relations will be discussed in a separate paper by Shahmoradi \& Nemiroff (2010{\it b}).
\\

Among all the gamma-ray observatories that have detected GRBs, BATSE provides the largest GRB database from a single experiment, consisting of observational data for 2704 GRBs. For many of the BATSE GRBs, high time and energy resolution data are available. The BATSE data are, therefore, the most suitable both in quantity and quality for detailed spectral studies and determination of the distribution of LGRBs on the planes of \sbep ~and \pbep, which can help to examine the veracity of the reported correlations.
\\

The plan of the paper is as follows: \S \ref{sec:BDT} is spent on the simulations of the BATSE Large Area Detectors (LAD) triggering thresholds and its possible effects on the bivariate distributions of GRBs in the planes of \sbep ~and \pbep. The possible reasons for the discrepancies between our results and the findings of N08 and G08 will also be given. We discuss the tightness of the Ghirlanda relation and its connection with the Amati relation in \S \ref{sec:Ghirlanda}. The properties of Amati relation in the observer and rest frames are investigated in \S \ref{sec:discussion}.  The results will be summarized in \S \ref{sec:conclusion}.

\section{BATSE DETECTION THRESHOLDS}
\label{sec:BDT}
\subsection{BATSE Trigger Algorithm}
\label{sec:BTA}
The investigation of the BATSE triggering threshold, its specifications as well as its possible effects on the detection of different types of GRBs, is well documented in articles by the BATSE team (e.g. Pendleton et al. 1995; Pendleton, Hakkila \& Meegan 1998; Paciesas et al. 1999; Meegan et al. 2000; Johnson, Meegan \& Hakkila 2000; Brainerd et al. 2000; Hakkila et al. 2003a; Hakkila et al. 2003b). Model-dependent studies of the BATSE triggering threshold for different $\epo$ -- based on the observational data as well as comparison with other detectors -- has also been presented by Band (2006), Band (2004), Band (2003) \& Band (2002). A precise determination of the trigger threshold, however, involves modeling the triggering algorithm of BATSE and its observational efficiency, taking into account the sensitivity of BATSE LADs at different angles of the incident photons. 
\\

BATSE had a relatively simple triggering algorithm compared to other instruments such as BAT onboard SWIFT and HETE-II (e.g. Band 2003). It generally triggered on a burst if the count rate in the second most brightly illuminated detector exceeded a threshold specified in units of standard deviations -- nominally $5.5\sigma$ for normal incidence on a single detector -- above background rate that was determined for each detector over a given time interval, usually set at 17.4 sec. However, as indicated by Band (2003), the requirement that a trigger occurs when the flux in the second most brightly illuminated detector reaches $5.5 \sigma$, raises the significance threshold from $5.5\sigma$ -- for a normal incidence -- to $5.96\sigma-7.78\sigma$ above the background, depending on the cosine of the angle between the normal axis of the detector and the source.  BATSE was programmed to trigger on any of three time scales: 64 ms, 256 ms and 1024 ms and the trigger energy range was generally set to 50 KeV - 300 KeV.
\\

To simulate BATSE, we use BATSE's Detector Response Matrices (DRMs) and public GRB data available through the online HEASARC archives \footnotemark \footnotetext{http://gammaray.msfc.nasa.gov/batse/grb/catalog/current/}. The fluences for four BATSE channels covering the energy range of $20-2000$ KeV -- which we treat as bolometric -- were taken from the ``Current BATSE" catalog. Jimenez et al. (2001) have shown that the fluences resulting from the processing pipeline used to create the BATSE catalog are slightly -- up to a factor of 2 -- larger than the fluences obtained by fitting high-resolution spectra. By simulation, using a wide range of $\epo$ ($1-2000$ KeV) with typical photon indices $\alpha = -1.1$ and $\beta = -2.3$, we find the ratio of the bolometric fluence (given the rest-frame energy range of $1-10,000$ KeV) as defined in G07 and the fluence in BATSE energy range to be very close to unity (on average $< 1.2$). Comparing this ratio with the results of Jimenez et al. (2001), extension of the energy band to compute bolometric fluence via spectral fits appears to be unnecessary for the current analysis (Band \& Preece 2005; Friedman \& Bloom 2005).

In some previous analyses (e.g. N08 \& G08), the trigger threshold limits were generally obtained using the Band model with typical photon indices fixed to $\alpha=-1.1$ \& $\beta=-2.3$ as a representation of the spectra of the whole sample of BATSE LGRBs. However, we will show in \S \ref{sec:sbep} \& \S \ref{sec:pbep} that these types of simulations are strongly affected by model biases and circularity problems. Instead, we rely on the light curves of the GRBs in their original forms as detected by BATSE detectors and use them to find the lower limits for the trigger threshold on \sbep ~\& \pbep ~planes by decreasing monotonically the amount of received photon counts in all energy channels for each individual burst. 

The analysis of background count rates in the light curves of 2145 BATSE GRBs gives average minimum required peak fluxes of 0.66-0.86, 0.34-0.43, 0.18-0.24 $ph\,cm^{-2}s^{-1}$ for the three timescales 64, 256 and 1024 ms respectively, given the aforementioned BATSE significance threshold. These values correspond well to the lowest observed peak fluxes of BATSE GRBs on these timescales. We also include all three timescales in the trigger criteria of our simulation. This is particularly important for SGRBs, as they are mainly triggered on 64 ms and 256 ms timescales. The entire simulation algorithms in this work are written in FORTRAN.

\begin{figure}
\includegraphics[scale=0.31]{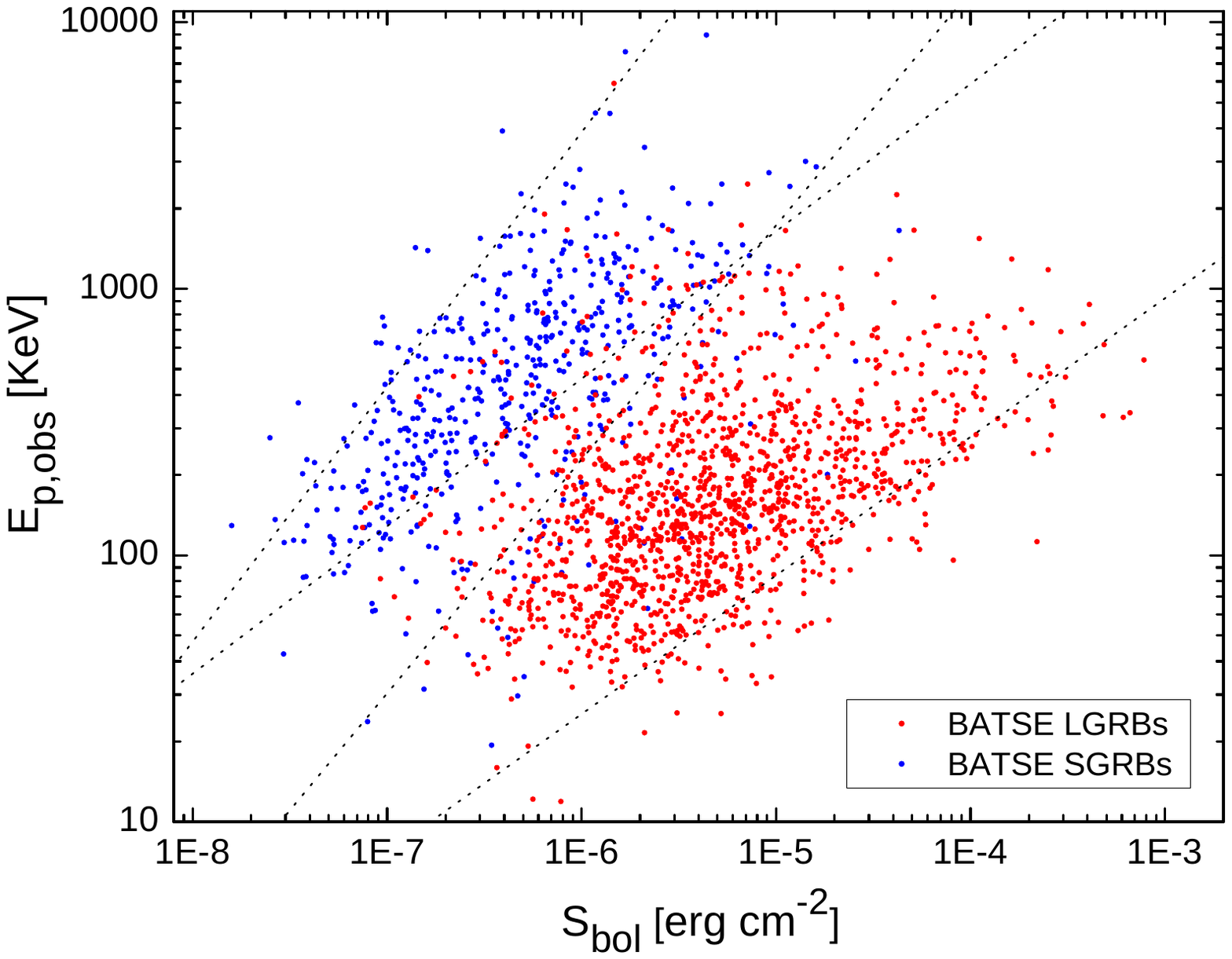}
\caption{Plot of the spectral peak energy, $\epo$, versus the bolometric fluence, $\sbol$, for 1900 BATSE GRBs with fluences, duration, and light curves available in the BATSE GRB catalog.  The red and blue dots represent LGRBs ($\tn > 3s$) and SGRBs ($\tn<3s$) respectively. The dotted lines are the approximate left and right boundaries for the distribution of each population (SGRBs \& LGRBs) obtained via a method described in \S \ref{sec:sbep}. Both SGRBs and LGRBs appear similarly bounded on both the upper left and lower right.
 \label{fig6}}
\end{figure}

\subsection{\sbep ~Plane of BATSE Bursts}
\label{sec:sbep}
Figure~\ref{fig6} shows the \sbep ~plane of 1900 BATSE bursts with continuous light curves and known durations and fluences, chosen from the complete BATSE catalog.  Interestingly, the distributions of both SGRBs (blue dots) \& LGRBs (red dots) look similar to each other and exhibit similar slopes on both their left and right sides (the dotted lines in Figure~\ref{fig6}).  The only difference is that SGRBs are shifted towards the upper left of the \sbep ~plane, possibly indicating that the duration of the burst has an important effect on its detection and determines the position of the GRB on the plots of $\epo$ versus the bolometric fluence ($\sbol$).  Moreover, the scatter of SGRBs on this plot appears to be smaller than the scatter of LGRBs. SGRBs also show a slightly tighter correlation (Kendall $\tau_{K}=0.43, 15\sigma$) compared to LGRBs ($\tau_{K}=0.32, 18\sigma$). This could be due to the narrower duration distribution of SGRBs as compared to LGRBs.

\begin{figure*}
\includegraphics[scale=0.26]{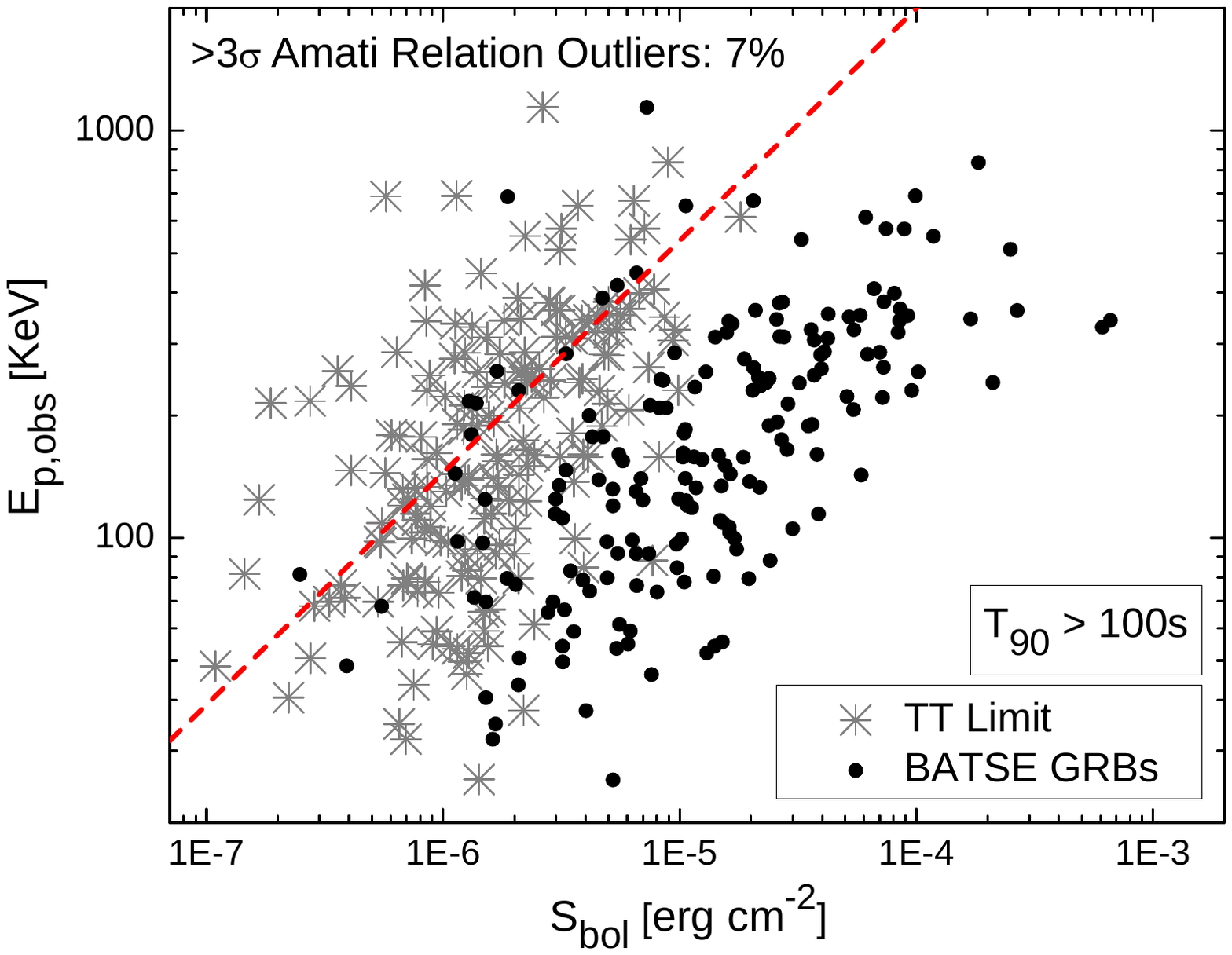}
\includegraphics[scale=0.26]{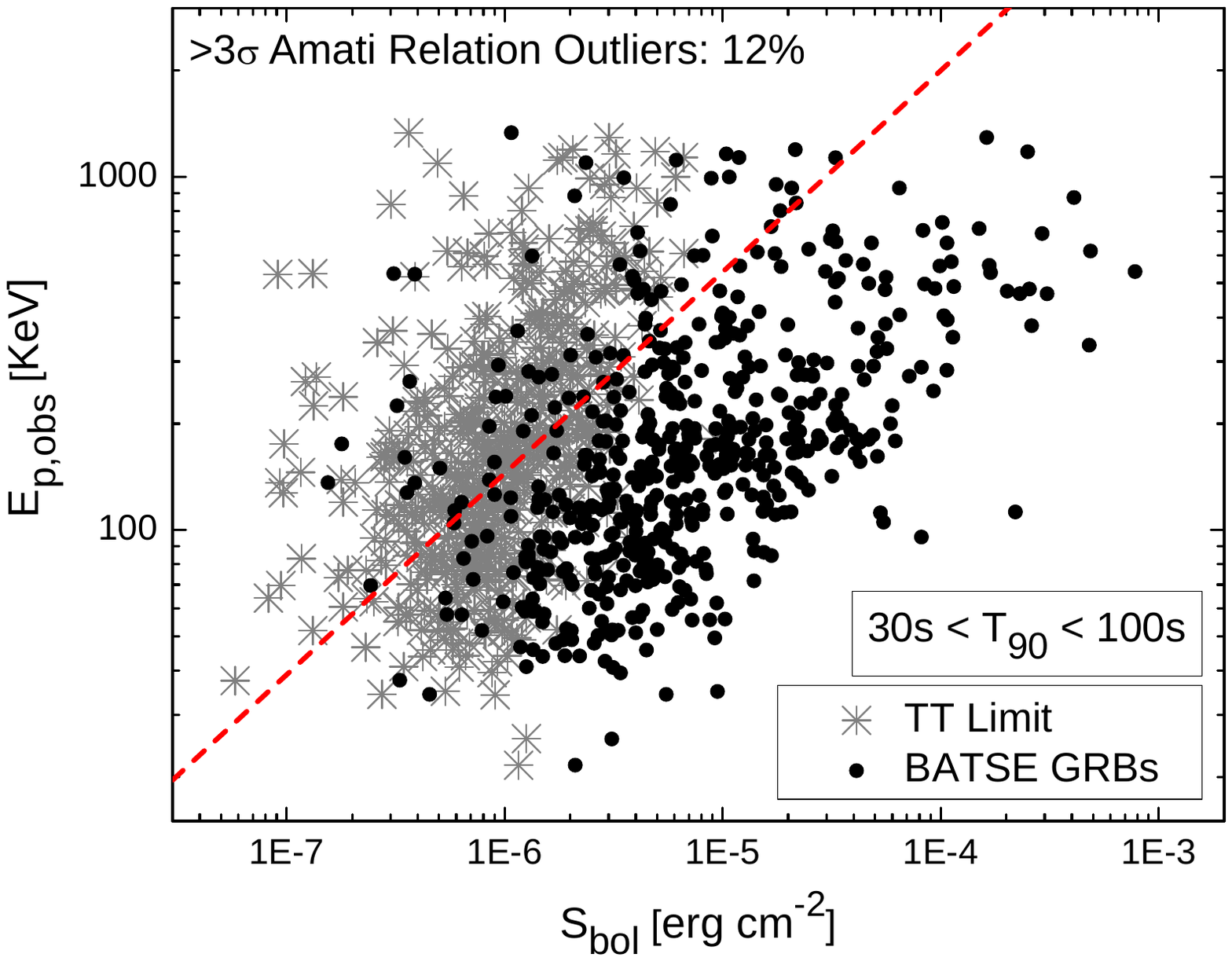}
\includegraphics[scale=0.26]{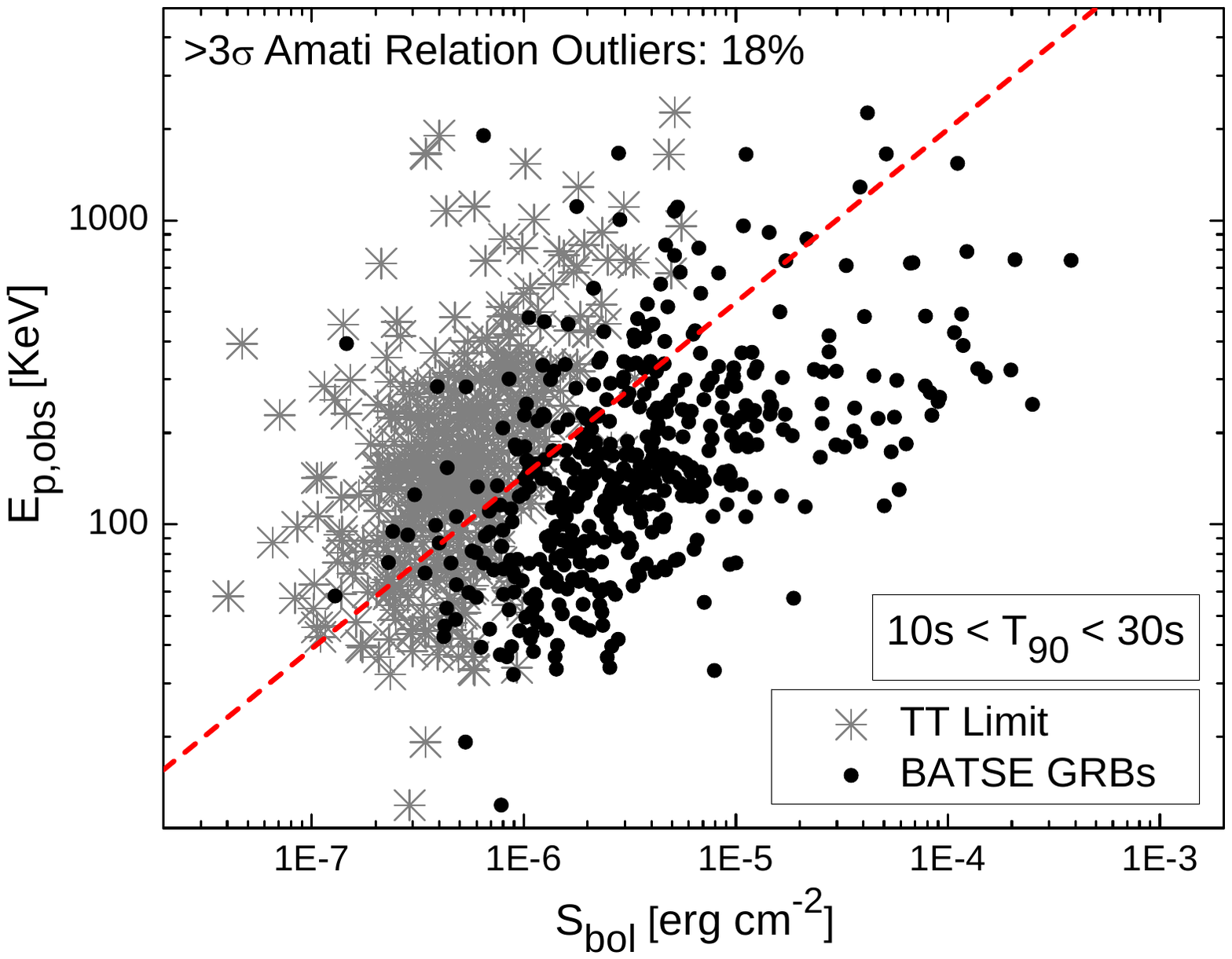}
\includegraphics[scale=0.26]{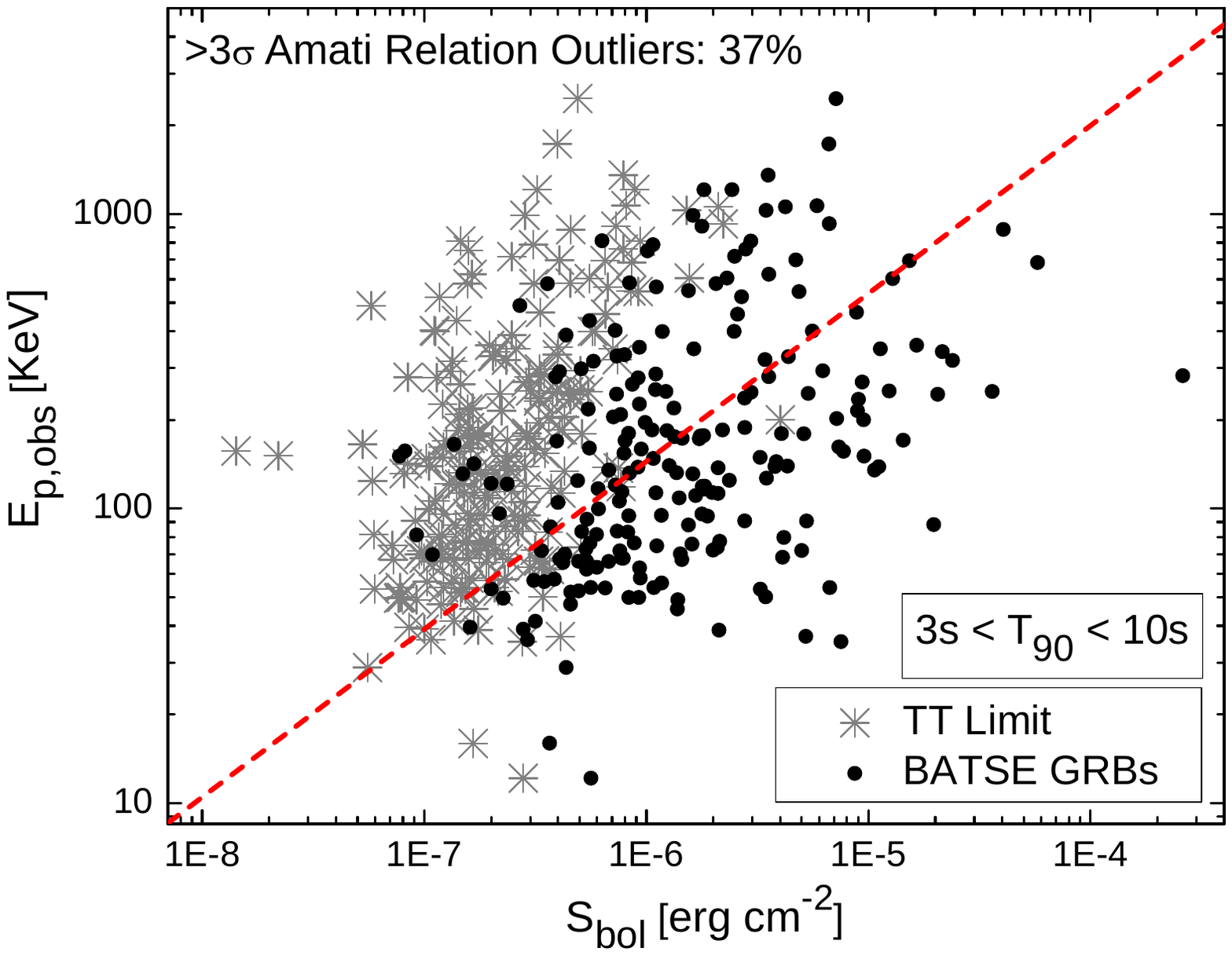}
\includegraphics[scale=0.26]{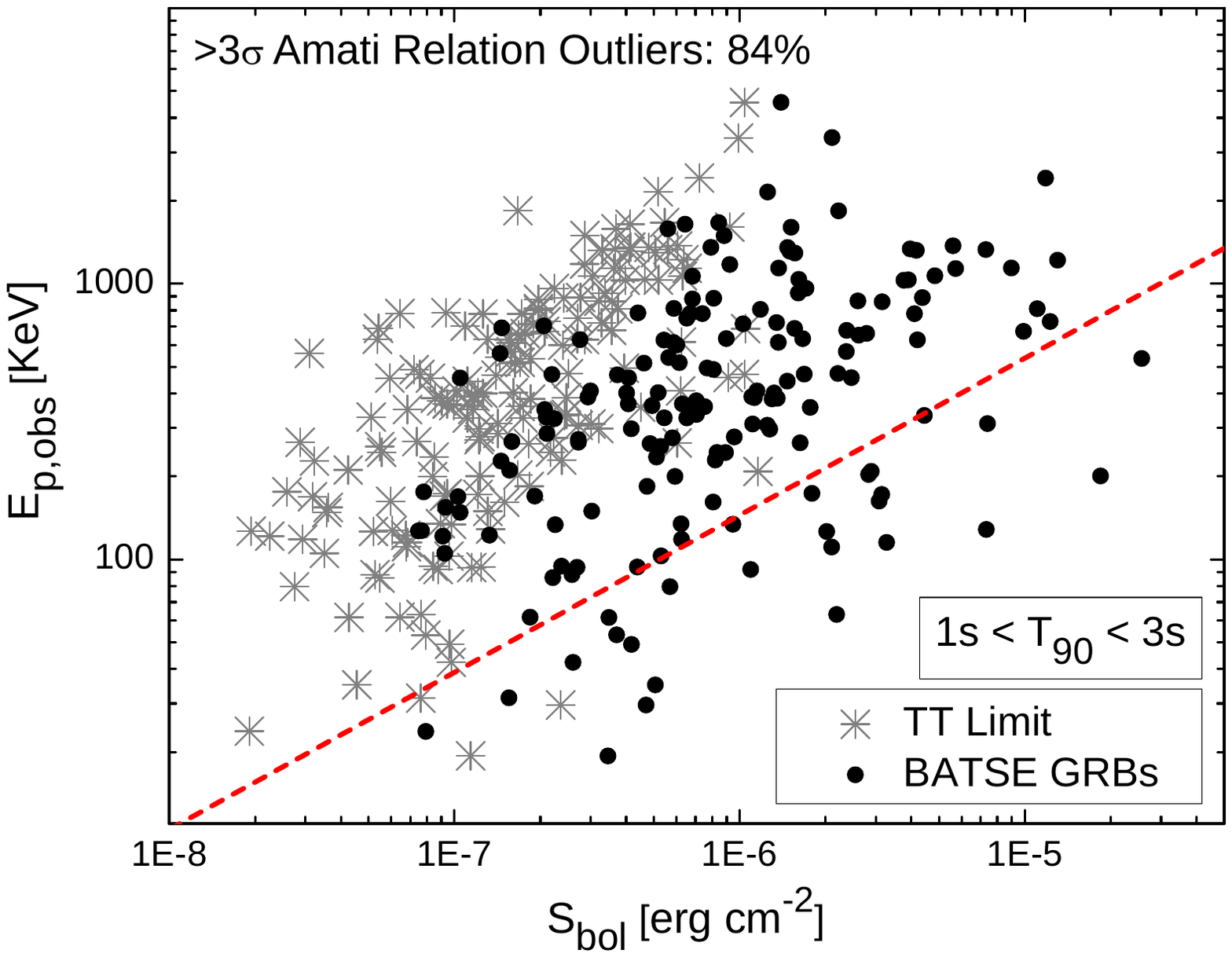}
\includegraphics[scale=0.26]{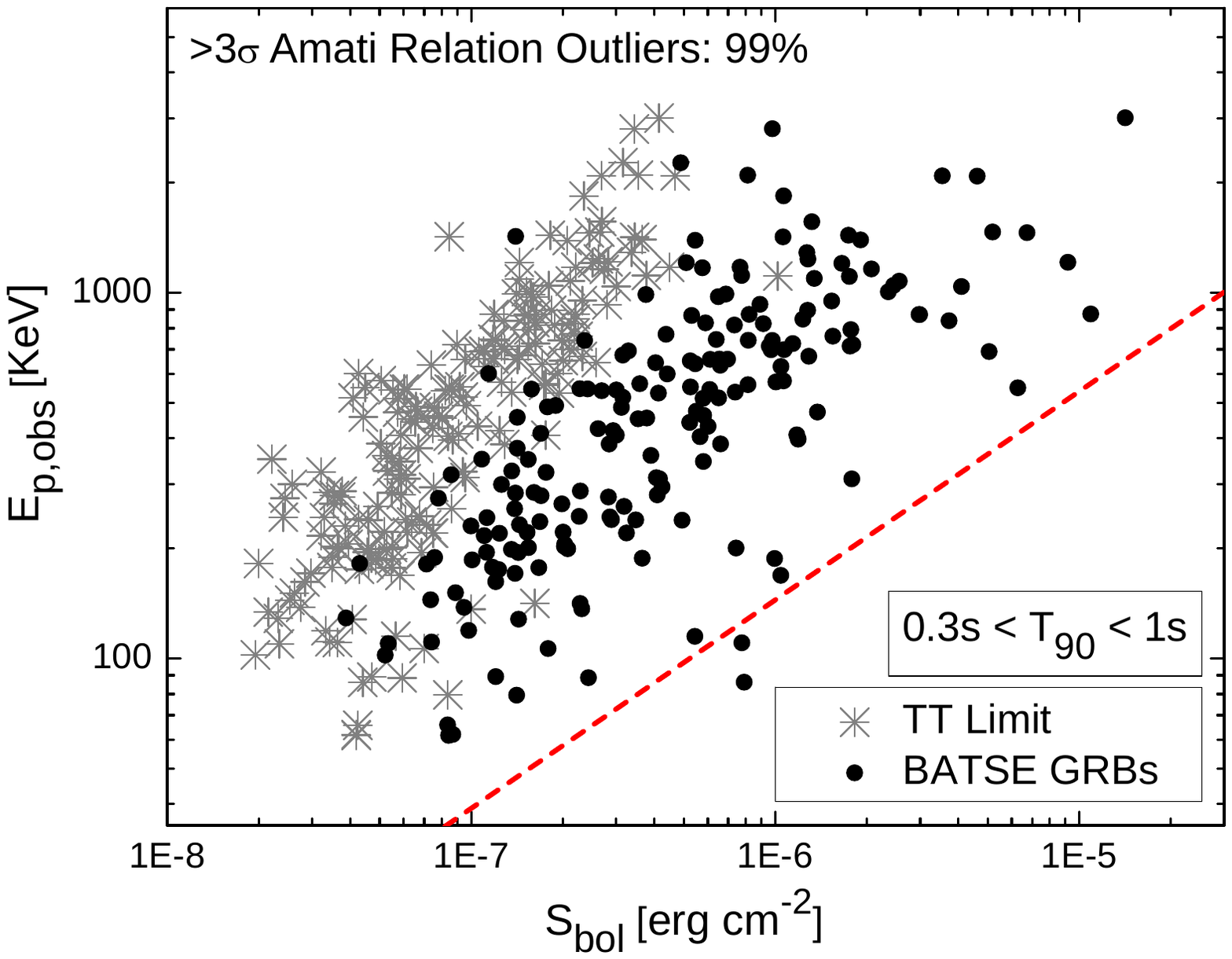}
\includegraphics[scale=0.26]{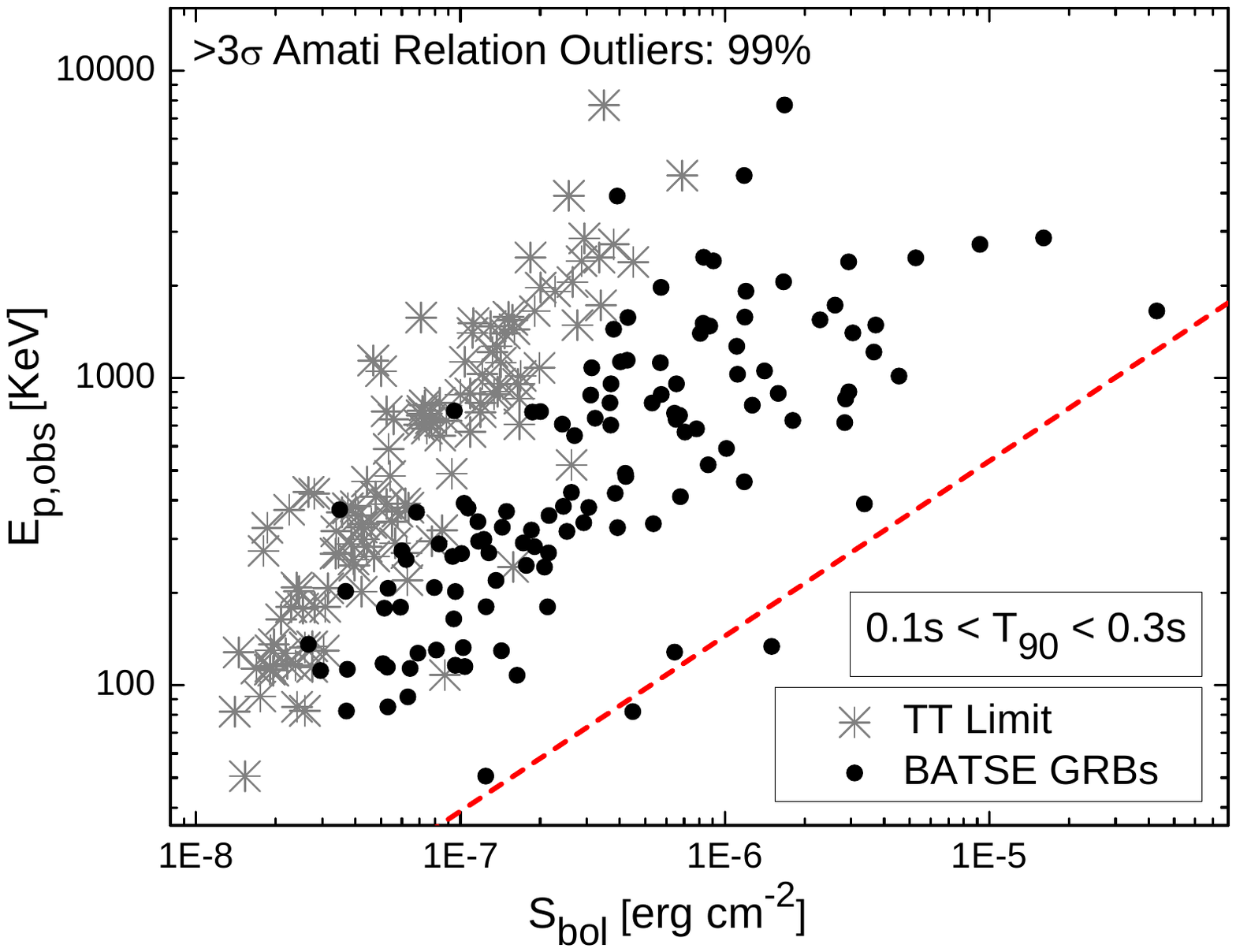}
\includegraphics[scale=0.26]{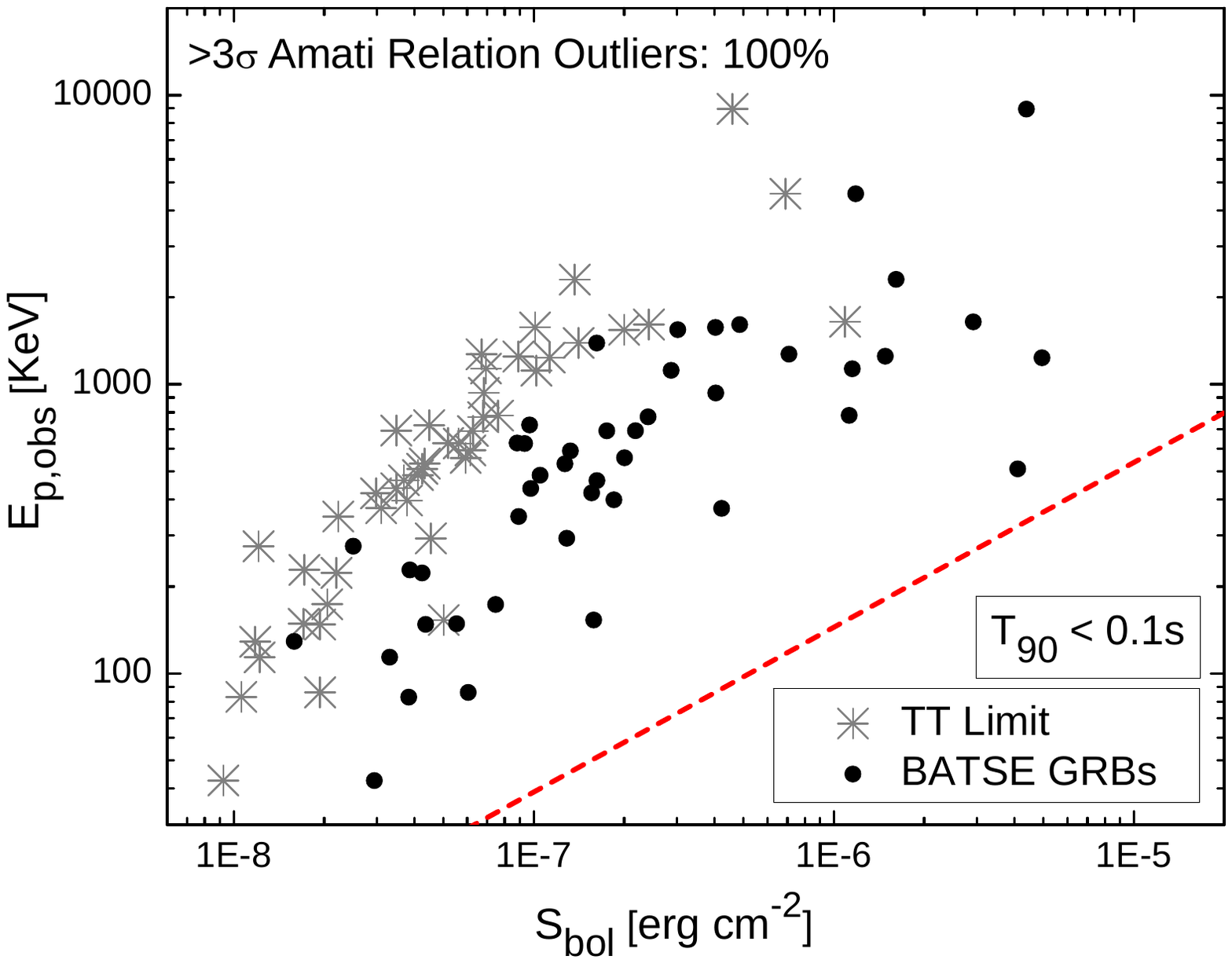}
\caption{Plots of $\epo$ versus $\sbol$ for different duration subsets of real BATSE (black points) and simulated BATSE (gray star) data.  Each simulated gray star represents the theoretically dimmest visible point for a particular BATSE GRB, below which BATSE would not have triggered on the burst, given BATSE detector thresholds. The dashed-red line in each plot represents the $3\sigma$ upper limit to the Amati relation (see \S3.2), above which any point is a certain $>3\sigma$ outlier to the Amati relation given by G07.\label{sbhrsim}}
\end{figure*}

\begin{figure}
\includegraphics[scale=0.31]{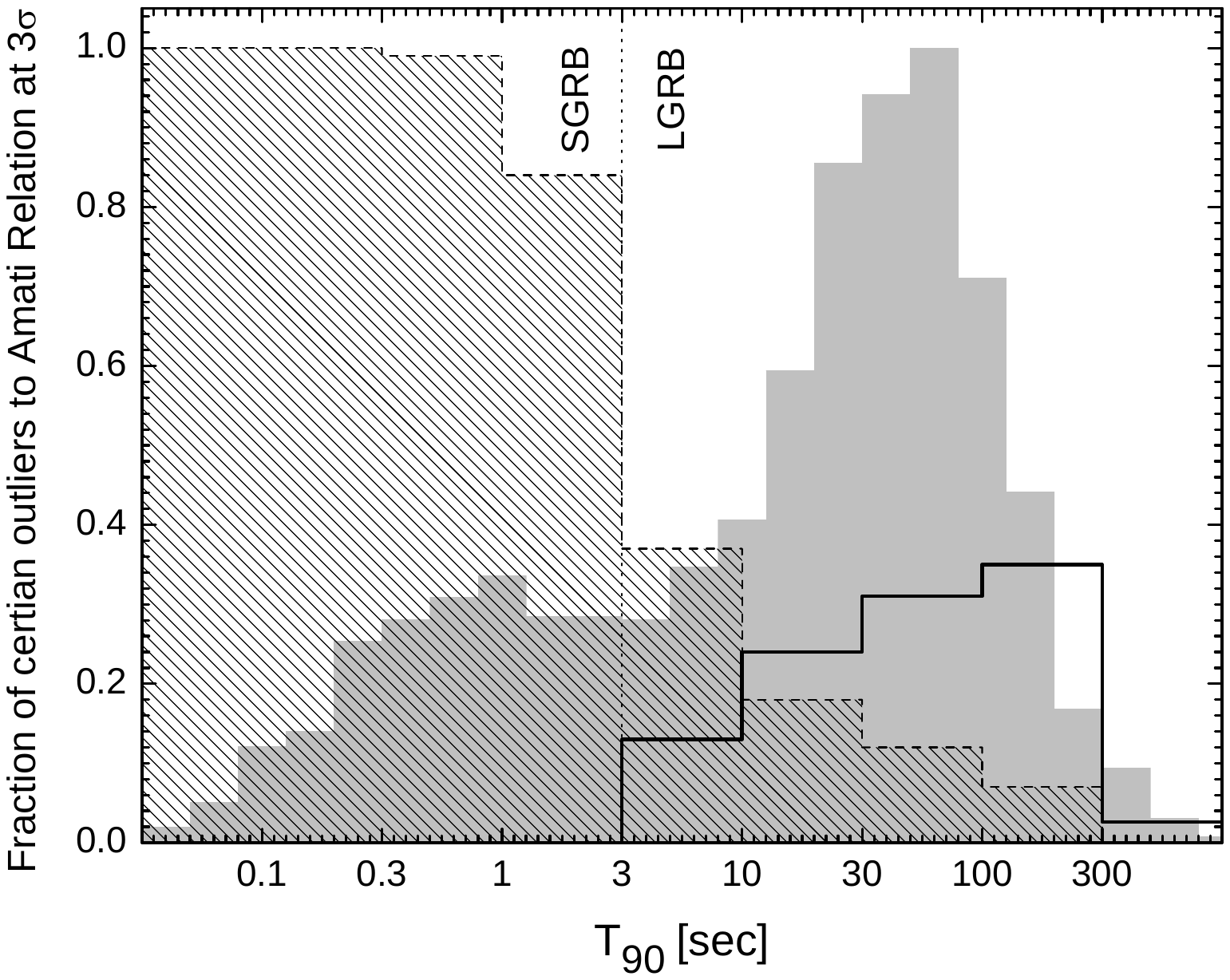}
\caption{Plot of the fraction of $>3\sigma$ outliers to the Amati relation versus $\tn$ duration for 1900 BATSE GRBs, represented by the hatched histogram.  A negative correlation is apparent, indicating that regardless of the type of the burst, whether LGRB or SGRB, the number of outliers to the Amati relation increases as the durations of the bursts decrease. The solid line represents the normalized histogram of $\tn$ of the LGRBs used by G08 to define the low-dispersion Log-Log linear Amati relation. As clear in the graph, most of the LGRBs in the G08 sample are chosen from a specific subsample of LGRBs that have very long durations. The background (gray) histogram represents the relative frequency histogram of $\tn$ for 2041 BATSE GRBs. Without recourse to any statistical significance tests, it is clear that {\bf the sample of GRBs used to define the Amati relation does not represent the entire population of Long-duration GRBs and it is biased towards the longest duration GRBs}. \label{fig7}}
\end{figure}
\begin{figure}
\includegraphics[scale=0.31]{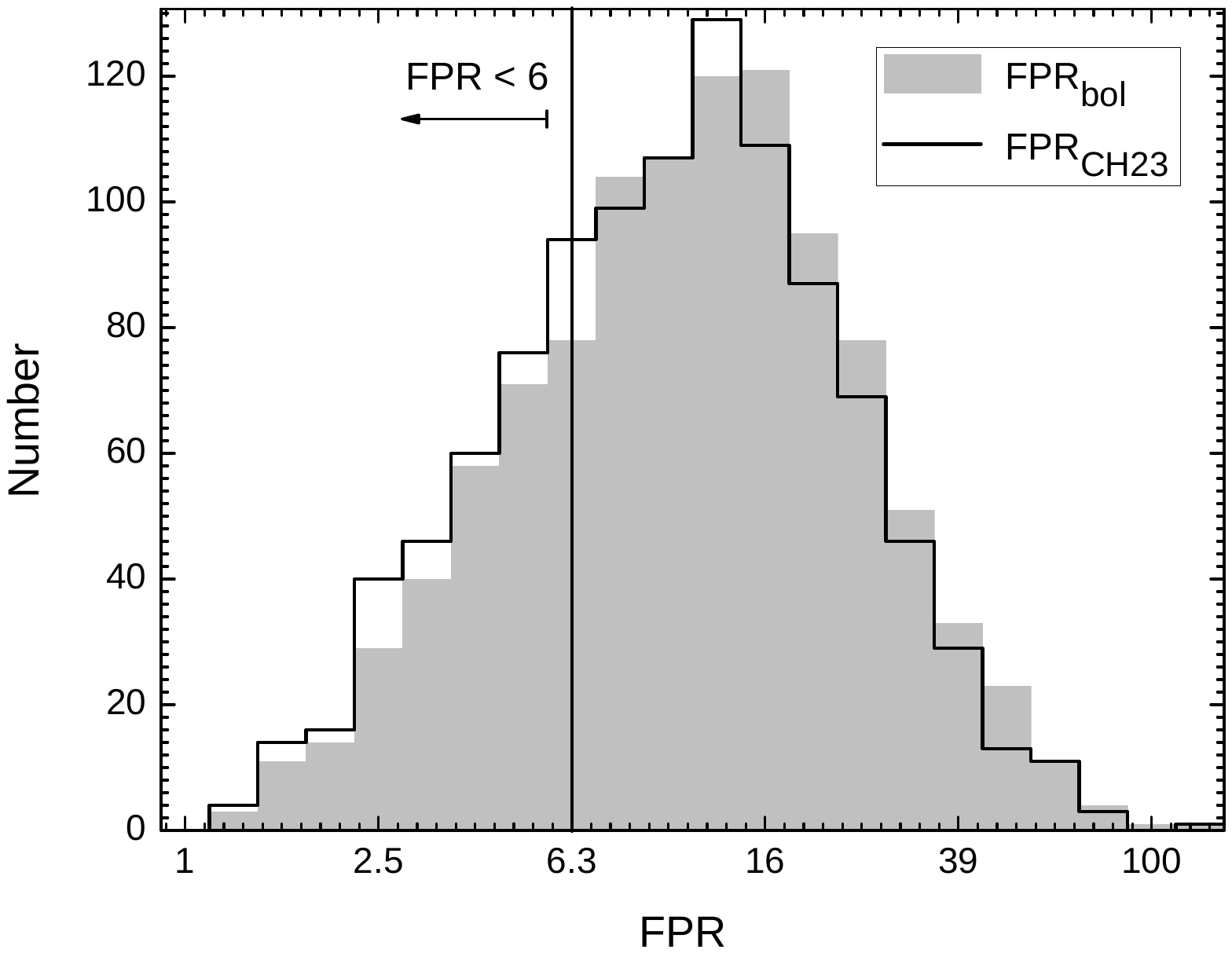}
\caption{Histogram of {\bf FPR}s, {\bf F}luence to 1-sec {\bf P}eak flux {\bf R}atios, for BATSE LGRBs.  Bolometric FPRs are plotted as the gray histogram, while the solid-line histogram includes FPRs calculated in the BATSE detection energy range 50-300 KeV. Both histograms peak at $FPR \approx 16$, in contrast to the peak value reported by G08 ($FPR \approx 6$) obtained for a smaller sample of BATSE LGRBs, represented by the vertical solid line in the graph. Only 27\% of the whole sample have $FPR < 6$. Underestimating this parameter can result in underestimation of the significance of BATSE detector thresholds on the distribution of LGRBs in \sbep ~plane, as explained in \ref{sec:sbep}.  This likely  explains one of the reasons for the significance of the differences between the results of the simulations of BATSE detector thresholds presented here (Figure~\ref{sbhrsim}) and the results obtained by other authors (e.g. Figures 1, 2, 5 \& 8 in N08 \& Figure 5 in G08). \label{FPRhist}}
\end{figure}

Since the duration of the burst plays an important role in its detection and also in connecting the triggering peak flux of the burst to its fluence, the effects of GRB duration should also be considered in the studies of selection effects on \sbep ~plane of GRBs.  To expand on this, we have created a simulation where the BATSE GRB sample is divided into eight subgroups with respect to their $\tn$ durations.  This simulation takes a given GRB and, while keeping duration constant, artificially decreases its fluence uniformly across the light curve until the GRB would no longer trigger BATSE.  The results of this simulation are shown in Figure~\ref{sbhrsim}. Each gray star shows the lowest fluence point for one particular BATSE GRB in the sample. Any point to the right of a gray star is detectable by BATSE {\textbf {\it if}} the burst has the same spectral parameters of the gray star, but a higher fluence.

One noticeable feature of the first four graphs on LGRBs in Figure~\ref{sbhrsim} is that the distributions of GRBs on each plane appears asymmetric along the principal axes of each of the distributions, implying that the left side of GRB distribution on each plot is likely truncated by the detector threshold.

Another important attribute of the plots in Figure~\ref{sbhrsim} is the negative correlation between the duration of the bursts and the fraction of `{\it certain}' outliers to the Amati relation at $>3\sigma$. The outliers are determined following the method described by NP05a and Band \& Preece (2005), also used earlier than these authors by Ghirlanda et al. 2004b for SGRBs. This negative correlation has possibly no physical origin and is due to selection effects caused by the durations of the bursts, that is, longer duration bursts are generally less detectable than shorter duration bursts of the same fluence. In other words, given the same fluence for two GRBs, the longer duration burst will have a larger {\bf F}luence to {\bf P}eak flux {\bf R}atio ({\bf FPR}) than the shorter duration burst which makes it less detectable for GRB detectors. 

The results of our simulations are clearly different from the findings of N08 \& G08 who obtain a distribution and slope for their sample of LGRBs that are far from BATSE trigger limits. The possible reasons for such discrepancy are summarized below:

\begin{figure}
\includegraphics[scale=0.31]{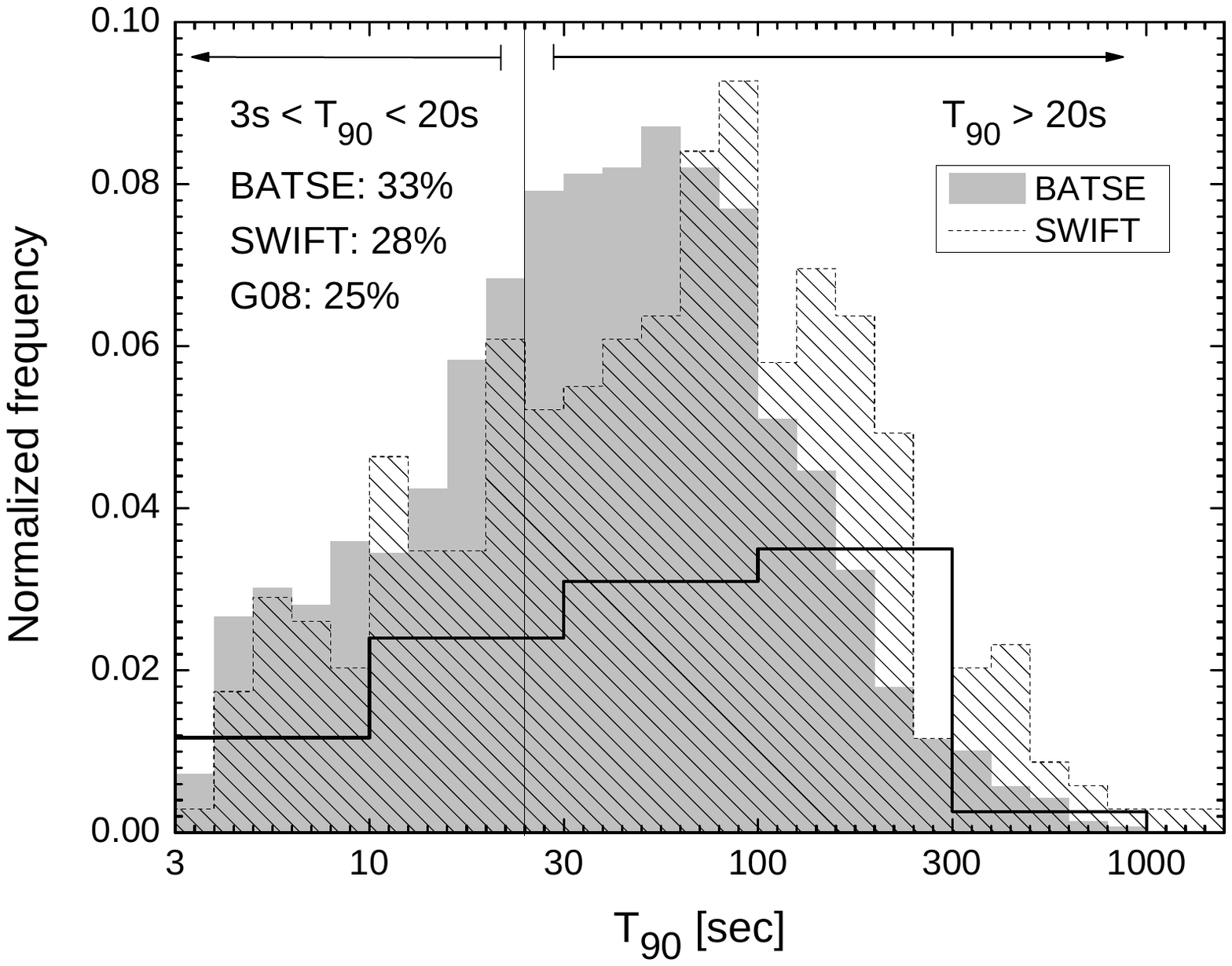}
\caption{A histogram of $\tn$ durations for 1390 BATSE LGRBs with $\tn>3 s$ (gray background histogram) as well as 345 SWIFT LGRBs with  $\tn>3 s$ (hatched foreground histogram).  Also shown is the $\tn$ histogram of 76 LGRBs used in G08 to construct the latest update of Amati relation, here normalized to 0.1 (rather than 1) for a better comparison of the three histograms. The solid vertical line represents $\tn = 20 s$ which is used by G08 \& N08 as the upper limit for the durations of the bursts in their simulations of Trigger and Spectral analysis Threshold (TT \& ST) limits.  However, only 33\% of the entire sample of BATSE LGRBs have $\tn < 20 s$. This ratio reduces to 28\% for the current SWIFT sample of LGRBs and to 25\% for G08 76 LGRBs used to construct the Amati relation. The latter is particularly important in the analyses done by G08, since they compare TT \& ST limits of different instruments obtained for $\tn < 20s$ with a sample of LGRBs that 75\% of them have $\tn > 20s$ that results in a significant underestimation of the selection effects. \label{durhist}}
\end{figure}

\begin{figure}
\includegraphics[scale=0.31]{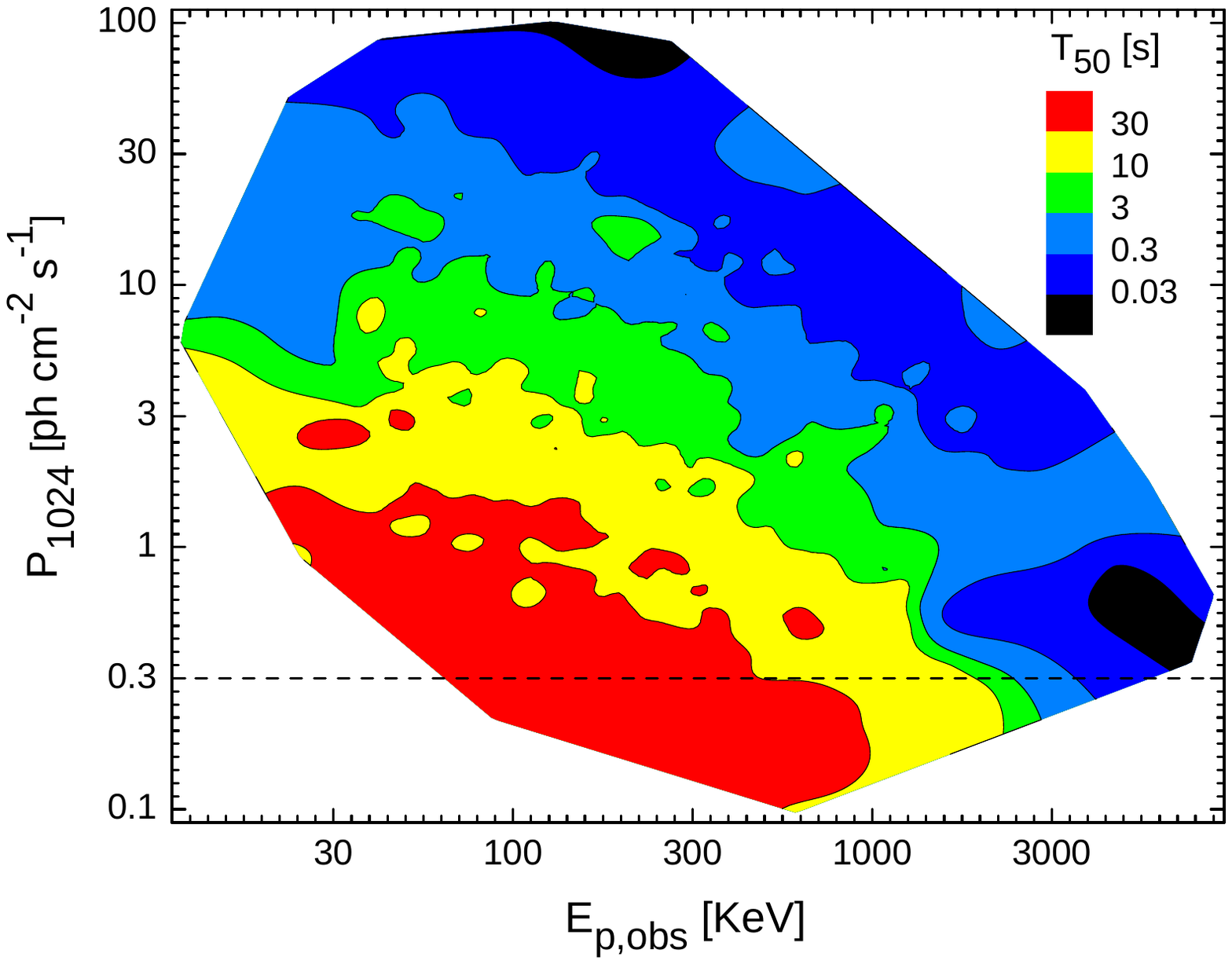}
\caption{A plot of the observed peak flux, $\pf$, versus the spectral peak energy, $\epo$, for 1900 BATSE GRBs segregated in $\tf$ by color. Each color region represents a population of the bursts with similar $\tf$ durations.  The fluences of all bursts are normalized to a fiducial fluence level ($10^{-5}$ $erg ~cm^{-2} s^{-1}$).  The vertical axis is the 1-sec peak flux of GRBs ($\pf$) and the dotted line represents the nominal BATSE trigger threshold on 1024 [s] timescale. For the same fluence and hardness (either $\epo$ or $\hr$), bursts with longer durations have peak fluxes that are orders of magnitude less than the peak fluxes of shorter duration bursts, and therefore will be less detectable.  This further illustrates the selection effect bias discussed in \S3.2. \label{fig8}}
\end{figure}

\begin{enumerate}
	\item {\bf Circular Logic Problem}: In the method used by N08 \& G08, the authors rely on the data from a fraction of LGRBs `{\it detected}' by BATSE and `{\it spectrally analyzed}' to constrain the parameters involved in their simulations.  These parameters, such as FPR, are required to relate the peak flux of the simulated burst to its fluence, and the duration of the burst. 

The use of the already detected GRBs to obtain the limiting parameters for their simulations, however, causes their analysis to suffer from a circular logic problem.   Specifically, the value of FPR that they use ($\approx\!6$), is representative of detected GRBs, not the entire population of GRBs which includes both detected and undetected bursts.  Figure~\ref{FPRhist} shows a frequency histogram of FPR values for 1053 bright BATSE LGRBs ($\tn\gtrsim3\,[sec]$) with nonzero fluences in all 4 energy channels.  Here FPR is defined as the ratio of fluence to {\it 1-second} peak flux ($\pf$).  Both fluence and the peak flux were calculated first in $50-300$ KeV energy range (the solid-line histogram of Figure~\ref{FPRhist}) and next for the whole BATSE energy range (the gray background histogram). The graph indicates a peak at FPR $\approx 16$, a factor of 2.5 higher than the FPR assumed in the simulations of N08 \& G08. The visible asymmetry in the histogram could be partly due to selection effects on high FPR LGRBs as they are generally less detectable compared to low FPR LGRBs.

	\item {\bf Duration Effects on GRB Detection}: Another factor, possibly responsible for the discrepancies between our results and the results obtained by N08 \& G09, is the range of durations of the bursts considered in the simulations.  N08 \& G08 use 5 seconds \& 20 seconds as the lower and upper limits. Such an assumption, however, underestimates the effect of threshold limits on very long duration GRBs that constitute a non-negligible fraction of the entire sample of LGRBs. In particular, for the BATSE sample of LGRBs considered in this work, the total number of bursts with $3s < \tn < 20s$ is only 33\% of the whole. This could be even lower, considering the fact that very long duration LGRBs have less chance to be detected. For LGRBs detected by the Burst Alert Telescope (BAT) onboard SWIFT satellite which is more sensitive to longest duration GRBs compared to BATSE LADs, this ratio decreases to 28\% (Figure~\ref{durhist}). 

In order to show the effects of duration on GRB triggering, we made a contour plot of $\epo - \pf - \tf$ for all BATSE GRBs by normalizing their fluences ($\sbol$) to a canonical fluence level ($10^{-5}$ $erg ~cm^{-2} s^{-1}$) in Figure~\ref{fig8}. Here we use $\tf$, since it is a more accurate measure of GRB duration than $\tn$ which is more dependent on the background fitting model and the choice of burst start and stop times, that are typically set by hand (e.g. Paciesas et al. 1999; B07). The graph clearly shows that for the same fluence and $\epo$, longer duration bursts have lower peak fluxes than the shorter bursts. In other words, given the same fluence \& duration for two bursts, the harder burst will have a lower chance of a triggered detection than the softer one. Were the above points considered in the analyses of N08 and G08, their {\bf T}rigger and {\bf S}pectral {\bf T}hreshold ({\bf TT} \& {\bf ST}) curves for BATSE would shift towards the distribution of BATSE LGRBs in their plots of \sbep ~(e.g. Figure~\ref{TTbol}). The two parameters -- FPR \& duration -- however, do not significantly affect the shapes of TT \& ST limits, since they only result in a shift in the positions of the entire TT \& ST limits on the plane of \sbep. The important factors in determining the shapes of these limits are discussed in \S \ref{sec:pbep}.
\end{enumerate}

   It is notable that two famous outliers of the Amati relation, GRB 980425 and GRB 031203, both lie in a region where the sample of G08 LGRBs exist and therefore cannot be flagged as certain $>3\sigma$ outliers to the Amati relation (G07) via the method given by NP05a and Band \& Preece (2005). This is a major weakness of the method given by these authors as indicated by themselves, since it can only set a lower limit for the number of certain outliers to the Amati and Ghirlanda relations. Nevertheless, following this method, for any linear relation among the rest-frame spectral parameters of GRBs -- $X_{rest}$ \& $Y_{rest}$ -- we can write,
\begin{equation}
\label{eq:rest}
Log(Y_{rest}) = \alpha + \beta Log(X_{rest}) + \xi \sigma.
\end{equation}
where $\xi$ stands for the significance level of being an outlier to a particular relation considered in the method. Here, we use $\xi=3$, corresponding to $3\sigma$ level (e.g. Figures~\ref{sbhrsim} \& \ref{outliers}), also a range of values for $\xi$ as depicted in Figure~\ref{outliers}. According to the assumed significance level $\xi$, Eqn.~\eqref{eq:rest} can then be transformed into the observer frame, knowing that,
\begin{eqnarray}
Y_{rest} &=& Y_{obs}f_{Y}(z) ~ ; ~ X_{rest} = X_{obs}f_{X}(z),\\
\label{eq:Ys}
\rightarrow \frac{Y_{obs}}{X_{obs}^{\beta}} &=& 10^{\alpha +\xi \sigma} \frac{[f_{X}(z)]^\beta}{f_{Y}(z)} = A(z,\alpha, \beta, \xi, \sigma).
\end{eqnarray}
where $f_{Y}(z)$ \& $f_{X}(z)$ are functions of redshift that relate the observer-frame spectral parameter to the corresponding rest-frame parameter. For the Amati, Ghirlanda, and Yonetoku relations, Eqn.~\eqref{eq:Ys} can be written as, 
\begin{eqnarray}
\label{eq:XYobs}
\frac{Y_{obs}}{X_{obs}^{\beta}} &=& 10^{\alpha +\xi \sigma} \left( 4\pi d_{L}^{2} \right) ^\beta \left( \frac{1}{1+z} \right) ^{1+\zeta \beta} \\ \nonumber
&=& A(z,\alpha, \beta, \xi, \sigma), \\
\zeta &=& \begin{cases} 1 & \text{Amati \& Ghirlanda relations} \\
						0 & \text{Yonetoku relation}\\
			\end{cases} \\
X_{obs} &=& \begin{cases} S_{bol} & \text{Amati relations} \\
						  P_{bol} & \text{Yonetoku relation}\\
						  f_{b}S_{bol} & \text{Ghirlanda relation}\\
			\end{cases}\\
Y_{obs} &=& \epo \\
d_{L} &=& \frac{c}{H_{0}}(1+z)\int^{z}_{0}dz'\big[(1+z')^{3}\Omega_{M}+\Omega_{\Lambda}\big]^{-1/2}
\end{eqnarray}

where $f_{b}$ is the beaming factor which is determined observationally from modeling the evolution of the afterglow for each individual GRB, $d_{L}$ is the luminosity distance in concordance cosmology assuming $\Omega_{M}=0.27$, $\Omega_{\Lambda}=0.73$, $H_{0}=72$ Km$/$sMpc (Komatsu et al. 2010) \& $c$ as the speed of light.

For the Amati and Ghirlanda relations, the function $A(z,\alpha,\beta,\xi,\sigma)$ has a maximum, $A_{max}(\alpha, \beta, \xi, \sigma)$. Any burst with spectral parameters such that,
\begin{equation}
\frac{Y_{obs}}{X_{obs}^{\beta}} > A_{max}(\alpha, \beta, \xi, \sigma),
\end{equation} 
will be a certain outlier to these relations at $>\xi \sigma$ for any redshift it might have. Using the above method we find that {\bf at least 19\% (or 21\%) of BATSE LGRBs are certain outliers to the Amati relation as given by Ghirlanda et al. (2008) -- (G08) -- at $>3\sigma$ significance level, based on the classical definition -- $\tn>3 ~[sec]$ -- (or fuzzy clustering classification) of BATSE LGRBs. Moreover, the consistency of the BATSE LGRBs with the Amati relation of G08 is strongly rejected with KS significance probability of $10^{-230}$ (Figure~\ref{outliers})}. This fraction could be even higher knowing that the current sample of 1900 GRBs used here is only $2/3$ of the total number of GRBs detected by BATSE, while the remaining GRBs -- not presented here -- are generally bursts with very low fluence close to trigger threshold limits, excluding exceptional bursts such as those with data gaps in their light curves.

In addition, it can be shown that the Amati relation as given by G08 is heavy on the side corresponding to the region of dim, hard bursts. Assuming a Gaussian distribution of the data around the best linear fit as considered by G07, one would expect to observe 50\%, 16\% and 0.023\% of the bursts lying at $>\xi \sigma$ for $\xi=0,1,2$ respectively, on the dim-hard side of the rest-frame Amati relation. However, the observed fractions are 80\%, 60\% \& 37\%, resulting in an observed excess of the bursts $>30\%$, $>44\%$ \& $>35\%$ at $>\xi \sigma$, $\xi=0,1,2$ respectively on the dim-hard side of the Amati relation. The expected \& observed fractions are plotted for a continuous range of significance levels in Figure~\ref{outliers} ({\it Lower Left \& Lower Right}).

\subsection{\pbep ~Plane of BATSE Bursts}
\label{sec:pbep}
Several authors have reported a strong correlation between $\epi$ and the rest-frame isotropic peak luminosity ($L_{iso}$) of LGRBs (e.g. Ghirlanda et al. (2009), hereafter G09; S07; Ghirlanda et al. 2005b; Y04; Schaefer 2003). The relation was originally constructed from a handful of BeppoSAX and BATSE bursts with known redshift and was updated by S07 for a sample of 64 LGRBs detected mainly by SWIFT and HETE-II and most recently by G09. The study of selection effects for this relation however, is much older than the relation itself (e.g. Lee \& Petrosian 1996; Lloyd \& Petrosian, 1999; Lloyd, Petrosian \& Mallozzi 2000). 

\begin{figure*}
\includegraphics[scale=0.31]{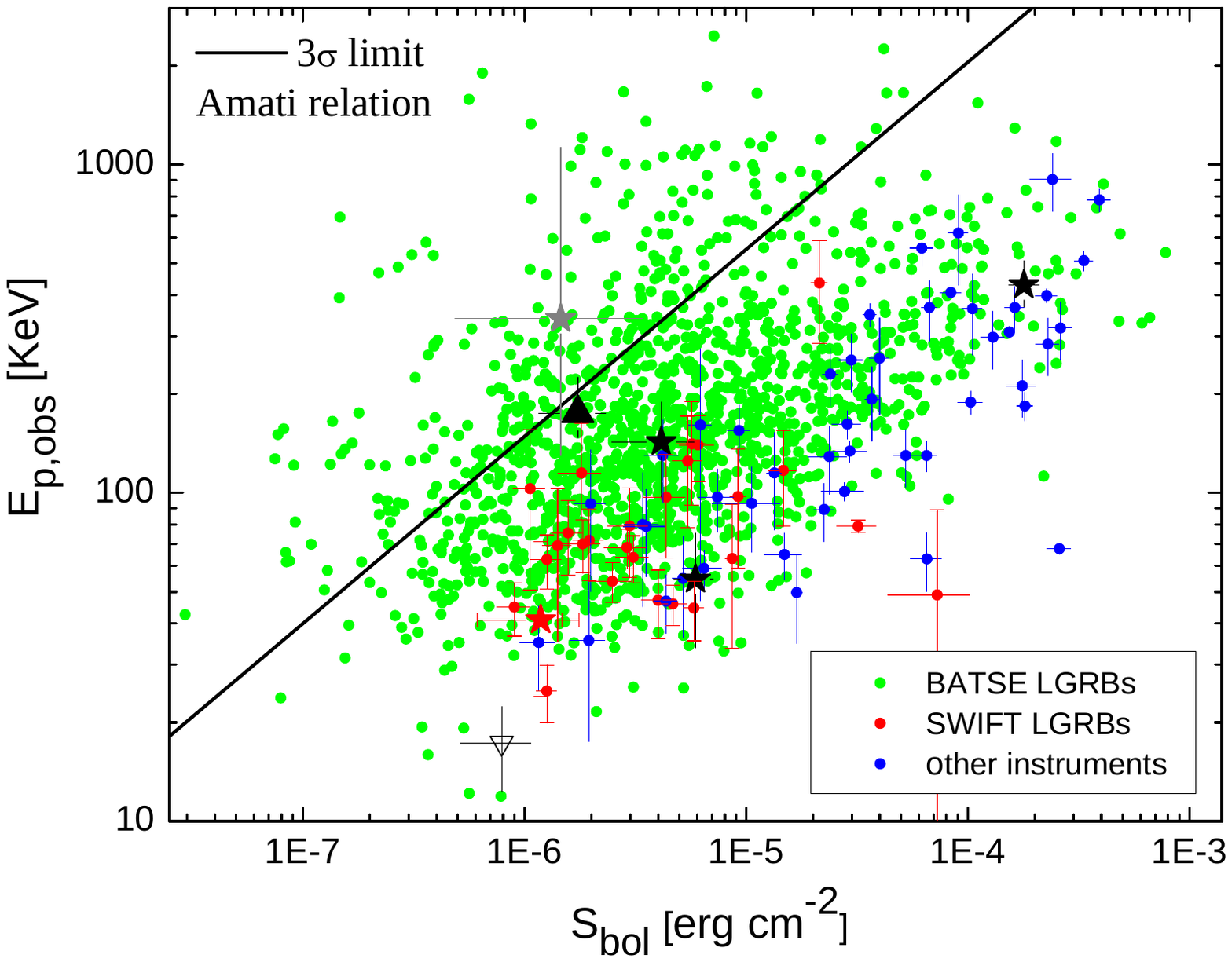}
\includegraphics[scale=0.31]{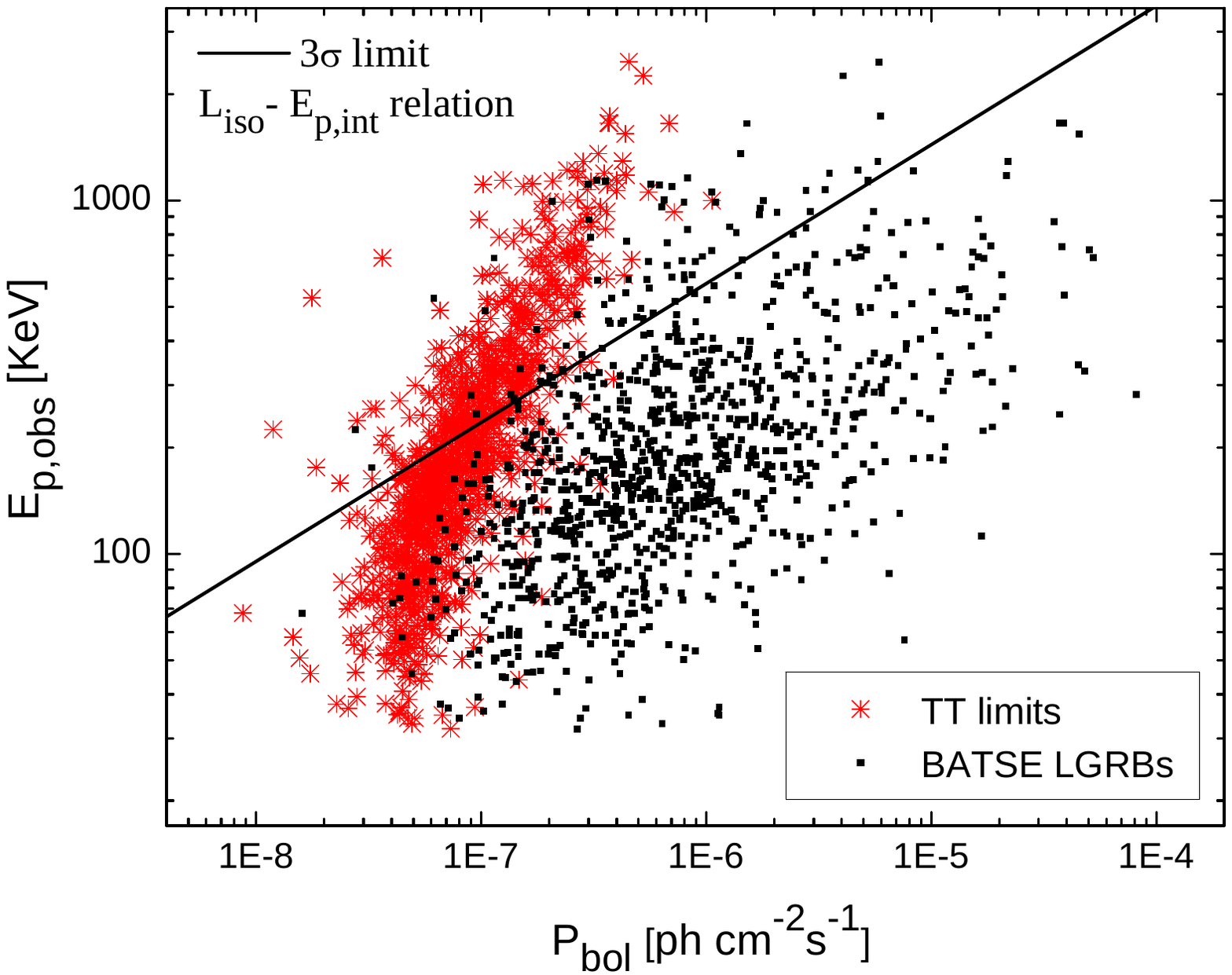}
\includegraphics[scale=0.31]{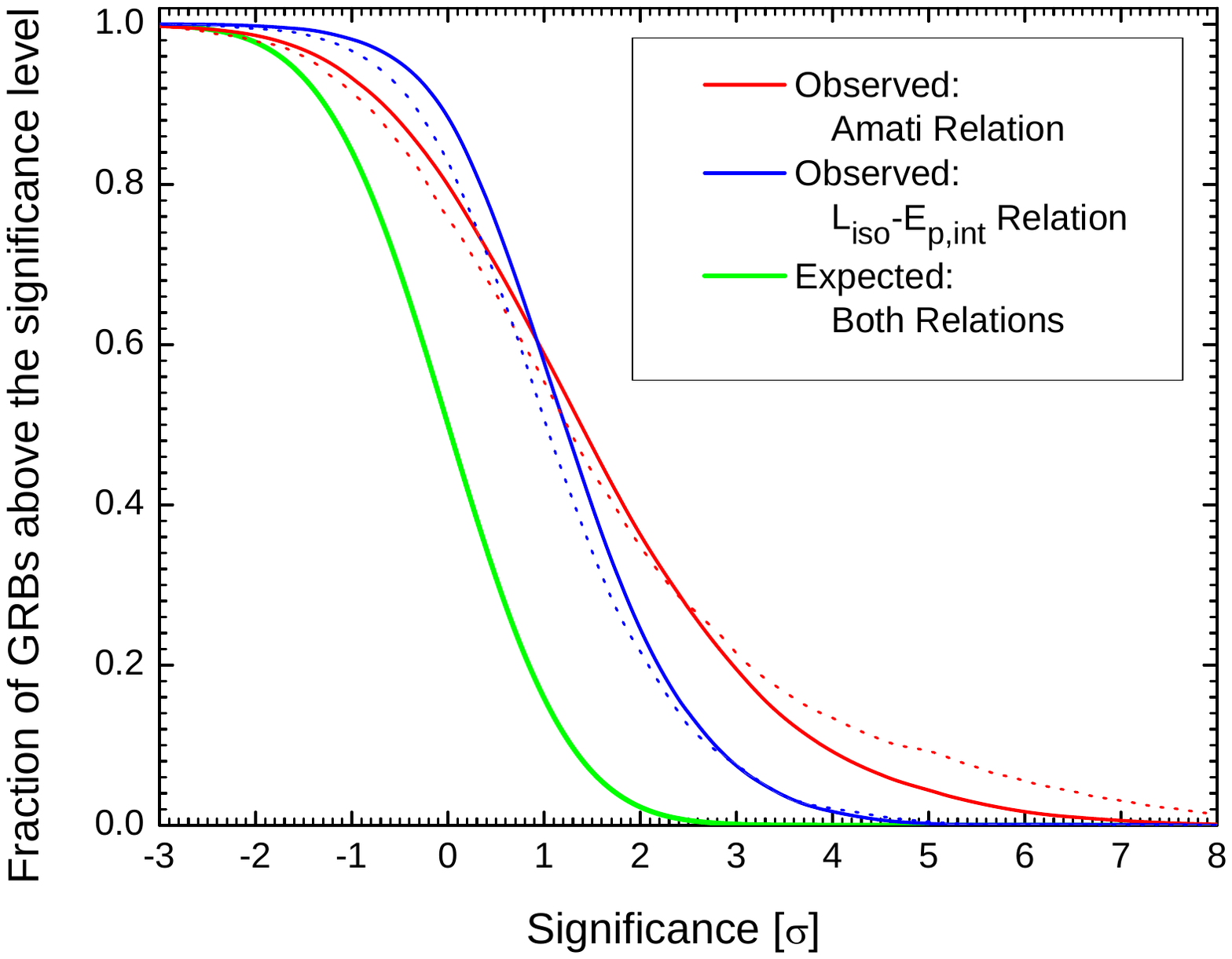}
\includegraphics[scale=0.31]{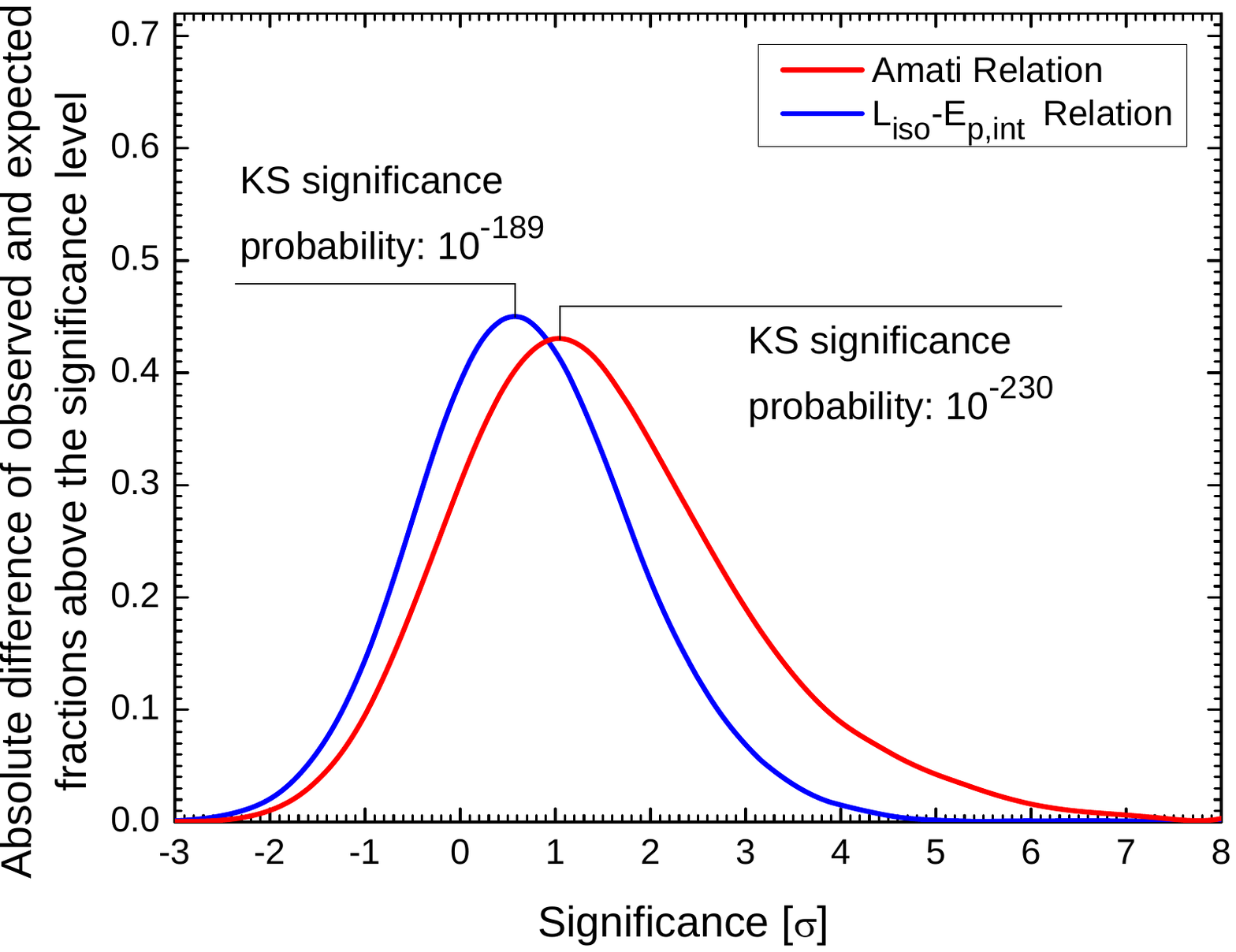}
\caption{{\it \bf Top Left}: Data plot of $\epo$ versus $\sbol$ for 1426 BATSE LGRBs among all 1900 BATSE GRBs, classified via fuzzy clustering based on their total fluence, hardness and duration, as well as the LGRBs of G08 used therein to construct the Amati relation. Green, red \& blue dots represent LGRBs detected by BATSE, SWIFT and other instruments, respectively.  The solid line represent the $3\sigma$ upper limit for the Amati relation as reported by G07. The 3 black stars represent GRB 980425 \& GRB 031203 (outliers to both the Amati and Ghirlanda relations), also GRB 070125, consistent with the Amati relation but an outlier to Ghirlanda relation at $5\sigma$, as reported by Bellm et al. 2008.  The black filled triangle represents GRB 060505, a far outlier to Amati relation. The gray star represents the sub-luminous SWIFT GRB 050826 (B07), the only GRB that is an outlier to the Amati relation via the method given by NP05a, regardless of its redshift. The white head-down triangle represents XRF 050416A and its consistency with the Amati relation is discussed in \S \ref{sec:discussion}. The red star represents GRB 060218 for the spectral parameters given by B07. {\it \bf Top Right}: A plot of $\epo$ versus $\pbol$ for 1053 bright BATSE LGRBs with cataloged nonzero fluence in all BATSE energy channels.  The black dots represent real BATSE GRB data, while the red dots represent the same BATSE bursts artificially reduced in fluence to the BATSE trigger threshold level.  The solid line is the $3\sigma$ upper limit for $L_{iso}-\epi$ relation given by Ghirlanda et al. (2009). While there is seemingly no selection effect present on the right side, the far left side of the sample appears to be affected by the BATSE trigger threshold limits. {\it \bf Bottom Left}: Plot of the fraction of BATSE GRBs that are certain outliers to the Amati \& $\liso-\epi$ relations at the given significance levels, represented by the red \& blue curves respectively. The solid \& the dotted curves are based on the classical definition of LGRBs ($\tn\gtrsim3\,[sec]$) and the fuzzy clustering classification of 1900 BATSE GRBs (Shahmoradi 2010), respectively. The green curve represents the expected fractions for both relations assuming the Amati relation given by G08 \& $\liso-\epi$ relation given by G09. {\it \bf Bottom Right}: The absolute differences between the observed and the expected fractions of BATSE LGRBs ($\tn>3\,[sec]$) -- represented as the solid red, blue \& green curves in the `{\it Lower Left}' plot -- at the given significance levels. {\bf Based on the traditional (or fuzzy clustering) classification of BATSE GRBs, at least 19\% (or 21\%) \& 8\% (or 8\%) of the BATSE sample of LGRBs considered in this work can be flagged as certain outliers to the Amati \& $L_{iso}-\epi$ relations respectively at $>3\sigma$ significance level. According to Kolmogorov-Smironov (KS) test, the similarity of the expected and the observed fractions for both the Amati \& $L_{iso}-\epi$ relations is rejected at high significance levels corresponding to KS significance probabilities of $10^{-230}$ \& $10^{-189}$ respectively.} Such extremely small probabilities invalidate the common assumption of Gaussian distribution of the residuals around the best fit Amati \& $\liso-\epi$ relations, at the best case, implying different normalization factors and significantly larger scatters for these relation than those proposed by Amati et al. (2006), G08 \& G09.  \label{outliers}}
\end{figure*}

Following the same method used for investigating the selection effects in the plane of \sbep ~in \S \ref{sec:sbep}, we use the light curves of BATSE LGRBs and their observed 1-second peak fluxes to study the bivariate distribution of GRBs in the \pbep ~plane.  This is done by gradually decreasing the normalization constant of the spectrum for each LGRB up to the limit where BATSE could not trigger the burst, based on the minimum required peak fluxes on the three different timescales that BATSE used for triggering: 64, 256 and 1024 ms. The results are shown in Figure~\ref{outliers} ({\it Top Right}). 

Similar to the \sbep ~plot (Figure~\ref{sbhrsim}), inspection of Figure~\ref{outliers} indicates that the far left side of the distribution of BATSE LGRBs on the plot of \pbep ~is affected by the trigger threshold.  Additionally, there appears to be an upper limit for $\epo$, given an observed fluence, on the far right side of the sample. Since maximum brightness is not limited at detection, this is unlikely to be due to data truncation (also depicted in Figure~\ref{pbcntepk}). Moreover, the left and right boundaries of the sample have different slopes: $m_{l} = 1.57\pm 0.05$ for $\epo \lesssim 400$ KeV \& $m_{l} = 0.89\pm 0.14$ for $\epo \gtrsim 400$ KeV, $m_{r} = 0.62\pm 0.02$. 

The GRB samples used by S07 \& G09 to construct the $L_{iso}-\epi$ relation, have correlation coefficients of $\tau_{K} = 0.48\pm0.07 (5.4\sigma)$ \& $\tau_{K} = 0.48\pm0.05 (6.7\sigma)$ respectively in the observer frame, while the sample of 1053 BATSE LGRBs considered here has $\tau_{K} = 0.36\pm0.02, 17\sigma$. Following the same procedure as for the Amati relation, we find that {\bf at least 8\% of BATSE LGRBs are outliers to $L_{iso}-\epi$ of G09 at $>3\sigma$ level for both the traditional ($\tn>3 ~[sec]$) or the fuzzy classification of BATSE LGRBs. In addition, the consistency of the BATSE LGRBs with $L_{iso}-\epi$ of G09 is strongly rejected with KS significance probability of $10^{-189}$ (Figure~\ref{outliers})}.

It can be shown that the $L_{iso}-\epi$ relation is heavy on the side corresponding to the region of dim, hard bursts. Assuming a Gaussian distribution of the data around the best linear fit as considered by N08, one would expect to observe 50\%, 16\% and 0.023\% of the bursts lying at $>\xi \sigma$, $\xi=0,1,2$ respectively, on the left side of the rest-frame $L_{iso}-\epi$ relation. However, the observed fractions indicate an excess of the bursts $>40\%$, $>43\%$ \& $>21\%$ at $>\xi \sigma$, $\xi=0,1,2$ respectively on the dim-hard side the $L_{iso}-\epi$ relation.

The results of the simulations shown in Figure \ref{outliers} ({\it Top Right}) again indicate the contamination of the BATSE LGRB distribution on \pbep ~plane by trigger threshold limits.  This is again contrary to the findings of previous authors, in particular N08 who find that the trigger threshold (TT) and spectral threshold (ST) limit curves for BATSE, as well as other instruments, are quite far from the bivariate distribution of the LGRBs on \sbep ~\& \pbep ~plots. The reasons for the discrepancies are likely as follows:

In `{\it model-dependent}' studies of selection effects it is usually assumed that all types of LGRBs with different $\epo$, have the same low and high energy photon indices on average, typically $\alpha ~ -1.1$ \& $\beta ~ -2.3$.  In these studies, the trigger as well as the spectral analysis threshold limits in \sbep ~\& \pbep ~planes are determined via a presumption that all LGRBs can be well described by a typical spectral model -- usually the Band model -- such as those done by G08 \& N08. This is true when the high- and low-energy photon indices are statistically independent of $\epo$. G08 find no dependency of $\alpha$ to $\epo$ for a sample of LGRBs used to construct the Amati relation therein. 

However, Shahmoradi \& Nemiroff (2010{\it a}) have shown that there is likely significant -- and in some cases strong -- positive correlations among the high- \& low-energy photon indices of the three GRB models: Band, COMP (CPL) \& SBPL and $\epo$ of bright BATSE GRBs.

\begin{figure*}
\includegraphics[scale=0.31]{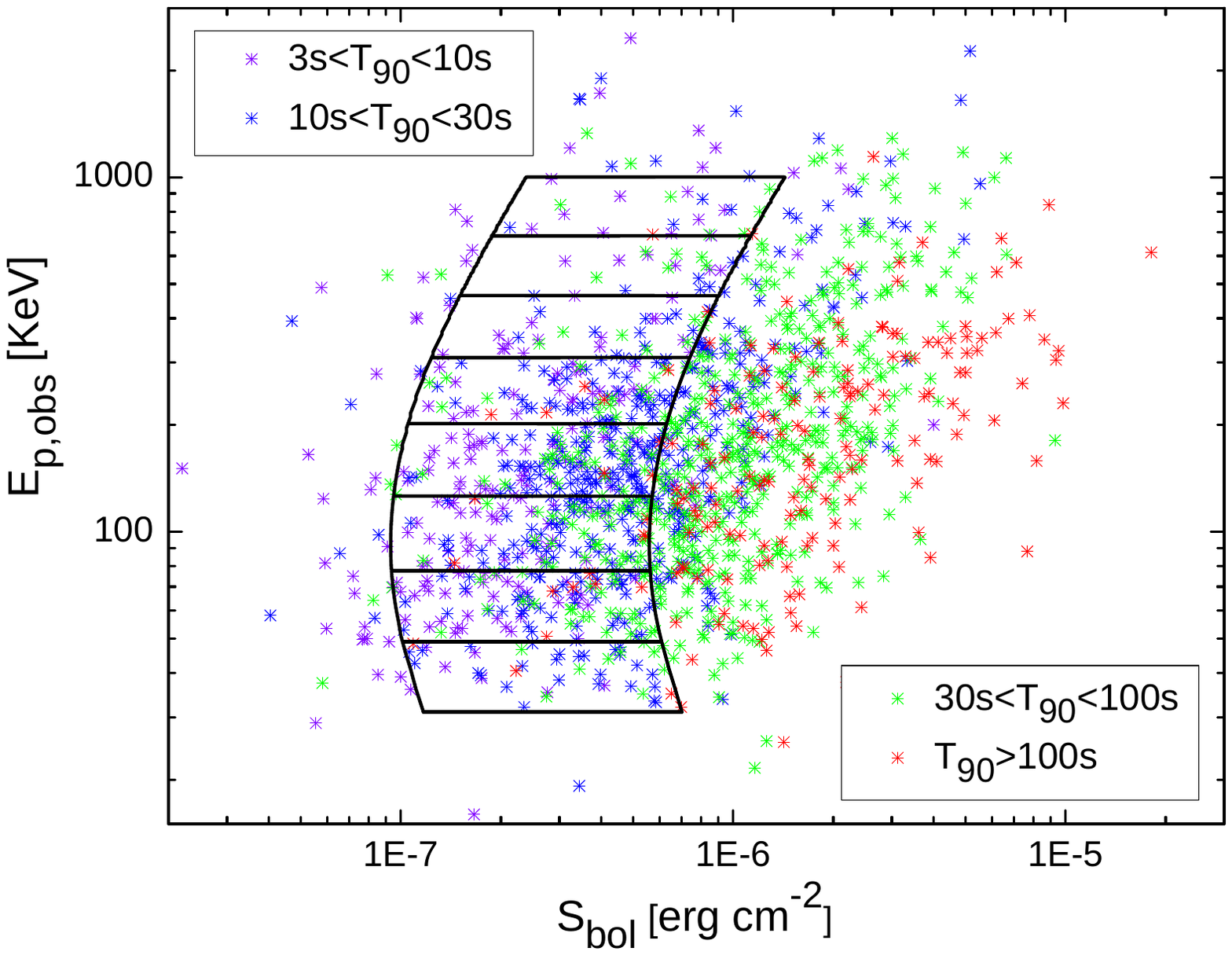}
\includegraphics[scale=0.31]{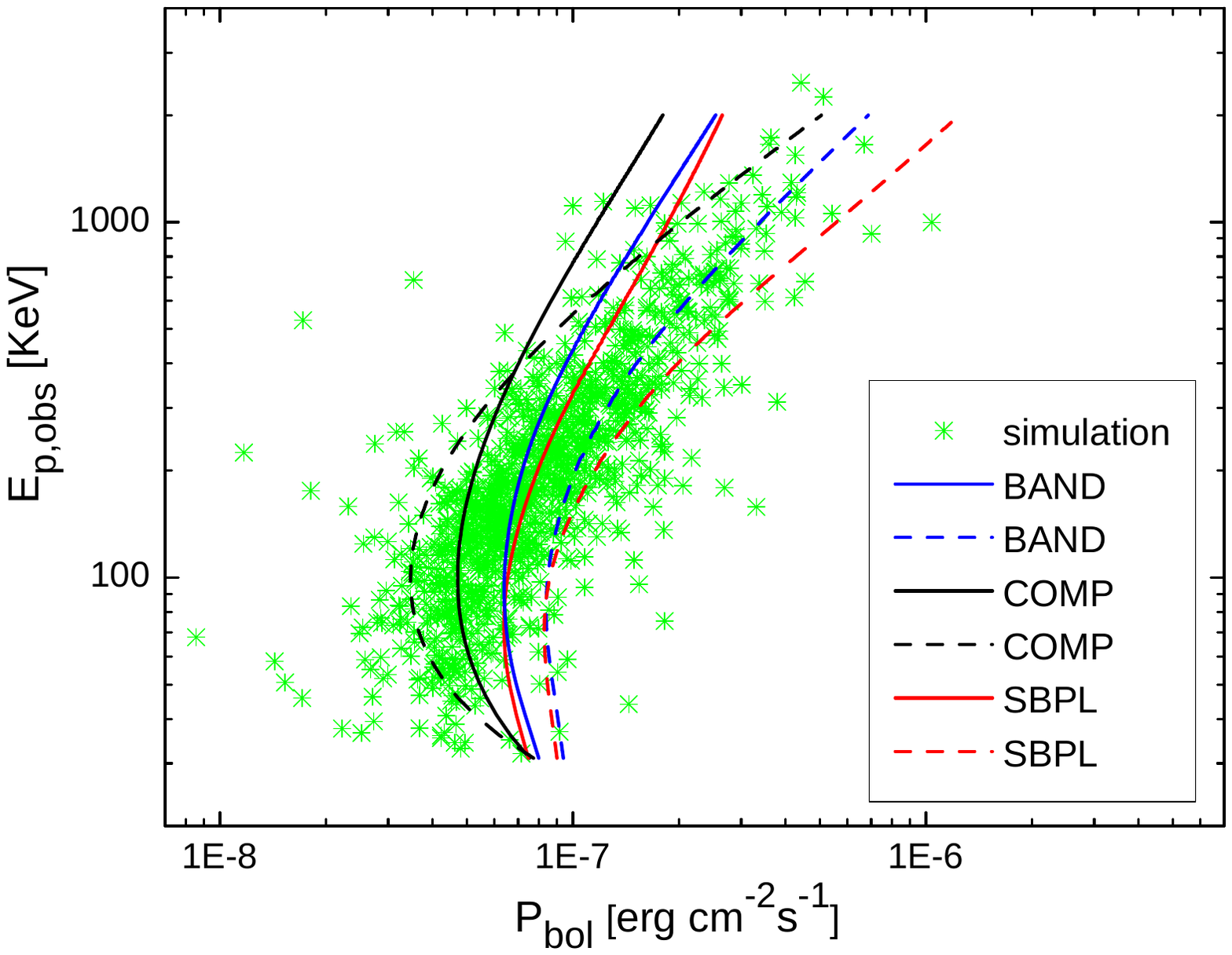}
\caption{{\bf Comparison of the simulated BATSE trigger limits based on the BATSE detected GRBs and the simulated `GRB-model dependent' BATSE trigger limits of G08 and N08}. {\bf Left}: Plot of $\epo$ versus $\sbol$ demonstrating trigger threshold effects for BATSE.  The colorful stars represent actual BATSE LGRBs ($\tn>3 s$) that are artificially dimmed in fluence to the detector threshold and so best show the real sensitivity of BATSE to dim LGRBs.  By contrast, the two superposed curves -- connected to each other by horizontal lines -- show the effective {\bf T}rigger {\bf T}hreshold (TT) limits obtained following G08 who use Band model with fixed average photon indices, $\alpha=-1.1$ \& $\beta=-2.3$, and FPR $=6$ in their analyses to estimate BATSE triggering threshold. The two curves correspond to the TT limits for $\tn=5$ sec (left) \& $\tn=20$ sec (right) as given in Figure 5 of G08. The plot indicates that G08 significantly underestimates the effects of BATSE trigger threshold on the distribution of BATSE LGRBs in \sbep ~plane (\S3.2). {\bf Right}: Plot of $\epo$ versus $\pbol$ demonstrating trigger threshold effects for BATSE.  The green stars represent 1053 actual BATSE bursts that are artificially dimmed in fluence to the detector threshold and so best show the real sensitivity of BATSE to dim GRBs.  By contrast, the superposed lines show the effective trigger threshold limits obtained using different spectral models.  The solid lines represent the TT limits for average photon indices of K06 sample of bursts with $\alpha=-1.1$ \& $\beta=-2.3$, $\lambda_{1}=-1.3$ \& $\lambda_{2}=-2.3$, $\alpha=-1.3$ for three spectral models Band, SBPL \& COMP respectively. The dashed lines are TT limits obtained using higher photon indices with $\alpha=-0.5$ \& $\beta=-2.0$, $\lambda_{1}=-0.5$ \& $\lambda_{2}=-2.0$, $\alpha=-0.5$ in the same order as mentioned. For the SBPL model, the fifth parameter (break scale) was fixed to the average value of K06 GRB sample ($\Lambda = 0.2$) in both cases.  None of the effective TT lines closely follow the actual TT points.  \label{TTbol}}
\end{figure*}

These positive correlations among the spectral parameters can have significant effects in {\it model-dependent} studies of selection effects. To illustrate this, we also simulated the BATSE trigger limits following the {\it model-dependent} methods as presented by G09, G08, N08 \& Band (2003). Figure~\ref{TTbol} ({\it Left}) is a plot of {\bf T}rigger {\bf T}hreshold (TT) curves assuming three different models (Band, SBPL \& COMP) for the GRBs with fixed photon indices, given in the caption of the figure.

The use of fixed averaged photon indices for all bursts with different $\epo$ clearly results in an underestimation of the minimum required bolometric peak flux ($\pbol$) to trigger a bursts at high $\epo$.  If this point is considered in `{\it model-dependent}' simulations of BATSE TT and ST curves, the slopes of  G08 \& N08 limiting curves would show much more proximity to the distributions of the bursts on \pbep ~\& \sbep ~plots. Among all the TT curves obtained for different spectral models, the COMP model curve with $\alpha=-0.5$ (black dashed line in the {\it Left} plot of Figure~\ref{TTbol}) best outlines the dimmest points of \pbep ~plane detectable by BATSE (green stars in the same figure).

\section{The Reality of the Ghirlanda Relation}
\label{sec:Ghirlanda}
Although the present and the previous analyses done by several authors (e.g. Butler et al. (2008) B07; NP05a; Band \& Preece 2005) strongly suggest that the Amati relation is due to complex selection effects not only in triggering process but also in the spectral analysis and redshift measurement, these cannot rule out other explanations for the inconsistencies.  Several alternative explanations were discussed by Ghisellini et al. (2006), the most popular being the off-axis model for GRBs that appear to be outliers to Amati relation. Therefore, as Amati (2008b) has suggested, the existence of low-dispersion -- as compared to Amati relation -- three-parameter correlations, such as the Ghirlanda relation and the empirical relation $\epi-\eiso-t_{jet}$ given by Liang \& Zhang (2005), could strongly favor a physical origin of the Amati relation and possibly other GRB correlations. To investigate such a possibility, we first determine the number of certain outliers to Ghirlanda relation for the homogeneous medium case as given by G07 under the extreme condition that all GRBs have a beaming factor $f_{b}=1$.  Following the same method used for the Amati \& $L_{iso}-\epi$ relations in the previous sections, we find only $<$0.6\% of the whole BATSE sample of LGRBs presented here to be certain outliers to the Ghirlanda relation (Figure~\ref{outliers}, {\it Top Left}). The fraction of outliers is therefore a significant difference between the Ghirlanda and Amati relations. In this section, we look into this difference more deeply. 

The Ghirlanda relation was initially presented by G04a for a sample of 16 LGRBs with jet opening angles estimated from the achromatic break of their afterglow light curves. They reported a significant scatter reduction and correlation improvement when the data were transformed form \eiep ~to \egep ~plane, with Spearman's rank correlation coefficients of $r_{s}=0.80$, $r_{s}=0.94$ and dispersions of 0.57 dex and $<$ 0.1 dex about the best fit lines of the Amati and Ghirlanda relations respectively. Interestingly, $\epi$ is highly correlated with both $\eiso$ \& $\eg$ in G04a sample. However, the hope for the existence of a significant difference between the correlation coefficients of the two relations vanishes when it is found that the Amati and Ghirlanda relations, {\it considering the same sample of bursts (i.e. 16 GRBs with measured jet opening angle given by G04a) for both relations}, have correlation coefficients that are within the $1\sigma$ uncertainties of each other.  The uncertainties in the correlation coefficients were determined via the bootstrap method by generating a large enough number of synthetic data sets.  The resulting Kendall rank correlation coefficients were $\tau_{K,A}=0.65\pm 0.16 ~(3.5\sigma)$ \& $\tau_{K,G}=0.85\pm 0.08 ~(4.6\sigma)$ for the Amati \& Ghirlanda relations respectively.  Moreover, the scatter of the Amati relation reduces to $\sigma_{A}=0.14$ when fit to the same sample of 16 LGRBs, which is comparable to the $\sigma_{G}=0.08$ for the Ghirlanda relation .  The apparent correlation difference between the two relations diminishes yet further when considering the whole sample of GRBs given in G04a, for which $\tau_{K,A}=0.74\pm 0.08 ~(5.2\sigma)$ \& $\tau_{K,G}=0.80\pm 0.07 ~(5.6\sigma)$. Both relations have the same scatter ($\sigma=0.15$) about their best linear fits. GRB 970508 was excluded from the above analysis because of its uncertain $\epo$ in G04a ranging from 145 KeV to $>$800 KeV,

The small size of the G04a sample with firmly reported $\theta_{jet}$ itself raises questions about the correlation improvement of the Ghirlanda relation.  Furthermore, were the cited correlation improvements to have a physical origin, they should manifest themselves more strongly in larger samples of GRBs. Unfortunately, a significantly larger sample is not yet available. 

A recent update of Amati, Ghirlanda \& Liang-Zhang relations have been given by G07, extending the number of GRBs with firmly measured spectral data from 16 in G04a to 24 in G07. Reanalyzing the sample of LGRBs given in G07, we confirm the correlations and scatters found therein: $\sigma_{A}=0.20$, $\sigma_{G}=0.09$, $\sigma_{LZ}=0.10$ ~for the Amati, Ghirlanda \& Liang-Zhang respectively. However, considering the same sample of bursts (i.e. only those with firmly measured spectral data, including $\theta_{jet}$) for all three relations, the scatter of Amati relation becomes comparable to the two others ($\sigma_{A}=0.14$).  Also, the correlation coefficient improvement observed in 16 GRBs of G04a sample deteriorates substantially: $\tau_{K,A}=0.76\pm 0.09 ~(5.2\sigma)$ \& $\tau_{K,G}=0.82\pm 0.06 ~(5.6\sigma)$.  Including GRB 070125, GRB 071010B \& GRB 050904, recently found outliers to the Ghirlanda relation at $>3\sigma$ level (Urata et al. 2009; Sugita et al. 2009; Bellm et al. 2008), makes the Ghirlanda relation comparable to the Amati relation and results in $\tau_{K,A}=0.76\pm 0.08 ~(5.5\sigma)$ \& $\tau_{K,G}=0.73\pm 0.07 ~(5.4\sigma)$ with $\sigma_{A}=0.16$, $\sigma_{G}=0.22$.

It is important to mention that we did not include any of the G07 bursts with uncertain spectral parameters (such as $\theta_{jet}$ or $\epo$) in the above analysis. The inclusion of all 33 bursts (by fixing the parameters with lower or upper limits to the values given) in G07 as well as GRB 070125, however, results in an even higher dispersion and lower correlation coefficient in the Ghirlanda relation as compared to the Amati relation: $\tau_{K,A}=0.67\pm 0.07 ~(5.6\sigma)$ \& $\tau_{K,G}=0.66\pm 0.07 ~(5.5\sigma)$ with $\sigma_{A}=0.18$  \& $\sigma_{G}=0.23$.

The same arguments in a yet stronger form hold for the comparison of the collimation-corrected peak luminosity ($L_{\gamma}$) with $\epi$ correlation and the $L_{iso}-\epi$ relation (Ghirlanda et al. 2005b, hereafter G05b). {\bf No significant difference in the scatter of the two $L_{\gamma}-\epi$ \& $L_{iso}-\epi$ relations is observed when we consider the same sample of 16 GRBs with firm jet opening angle -- reported in G04a -- for both relations (0.15 dex \& 0.16 dex respectively)}.

How important are the $<$0.6\% of certain outliers to the Ghirlanda relation?  Unlike the case for the Amati relation, the method given by NP05a to determine the number of certain outliers, cannot be regarded as a rigorous test of the Ghirlanda relation, since it requires a knowledge of the jet opening angle distribution function as well as the redshift distribution of the bursts. Several attempts have been made so far to determine the distribution of jet opening angles (e.g. Frail et al. 2001; Norris 2001; Ghirlanda et al. 2005a; Guetta et al. 2005; Friedman \& Bloom 2005), even given the sparse number of bursts with measured jet opening angles. However, as indicated by Band \& Preece (2005), also by Perna, Sari \& Frail 2003, a major problem with the current observed distribution of $\theta_{jet}$ is that it is affected by another type of selection effect on its head and tail (i.e. very high and very low $\theta_{jet}$), which is related to the current limited ability of observing very early and late breaks in the afterglow evolution of the bursts. Nevertheless, all obtained distribution functions imply a range of $\theta_{jet}<40^{\circ}$ for the jet opening angle with the peak of the distributions being around $5^{\circ}-10^{\circ}$. Therefore, the extreme assumption that we made at the beginning of this section (i.e. $f_{b}=1$) in order to determine the number of outliers to Ghirlanda relation appears to be unrealistically generous.  By this, we have assumed that the outflows of all bursts are essentially isotropic.  Even if the relativistic outflows are not highly collimated, some beaming is expected in most cases, since the energy channels mainly along the rotation axis of the inrushing material into the newly created black hole (Woosley, 1993).  

We therefore conclude that the use of $f_{b}=1$ in the method given by NP05a, severely underestimates the number of outliers to Ghirlanda relation. In order to show how the apparent consistency of the bursts with Ghirlanda relation exacerbates using $f_{b}<1$, we also find the Ghirlanda relation limit at $>3\sigma$ using the latest update of the relation (G07) and assuming $\theta_{jet} = 25^{\circ}$ as the upper value of the jet opening angle -- twice as large as the largest reported angle in the GRB sample of G07 -- for which we find 7\% inconsistency given the BATSE sample of LGRBs considered in this work. Another consistency check can be performed by estimating the beaming fractions ($f_{b}$) of bright BATSE LGRBs via the relation proposed by Norris (2001) between the spectral lag of LGRBs and their beaming fractions. Even though we make generous assumptions in the calculation of the number of certain outliers, we find at least $34\%$ of 1310 bright BATSE LGRBs to be certain $>3\sigma$ outliers to the Ghirlanda relation (Eqn. 3 of G07).

In addition, the redshift that maximizes the redshift-dependent term $A(z,\alpha, \beta, \xi, \sigma)$ of NP05a's method in equation \eqref{eq:XYobs} is at $z_{max}>10$ for the Ghirlanda relation. This redshift is much larger than the maximum detectable redshift by BATSE that Cohen \& Piran (1995) report ($z=2.1^{+1}_{-0.7}$) assuming no evolution in the luminosity function of the bursts. Moreover, Norris \& Gehrels (2008), have recently estimated the redshift distribution of SWIFT GRBs with unknown redshifts to be the same as the rest ($1/3$) of the SWIFT sample with measured redshifts resulting in an average $z\sim 2.1$ for the whole sample of SWIFT bursts. Knowing that BATSE LADs were on average 5 times less sensitive than BAT (Fenimore et al. 2004), we can use this as an upper limit for the average redshift of BATSE GRBs which results in an $A(z)$ that is a factor of two smaller than what was used to obtain the limits given in Figure~\ref{outliers} {\it Top Left}, leading to an increase in the number of certain outliers to Ghirlanda relation.

The consistency checks for this relation are, however, uncertain so long as the accurate unbiased redshift and jet angle distributions of the whole sample of BATSE bursts are not known.

\section{Discussion}
\label{sec:discussion}
\begin{figure*}
\includegraphics[scale=0.31]{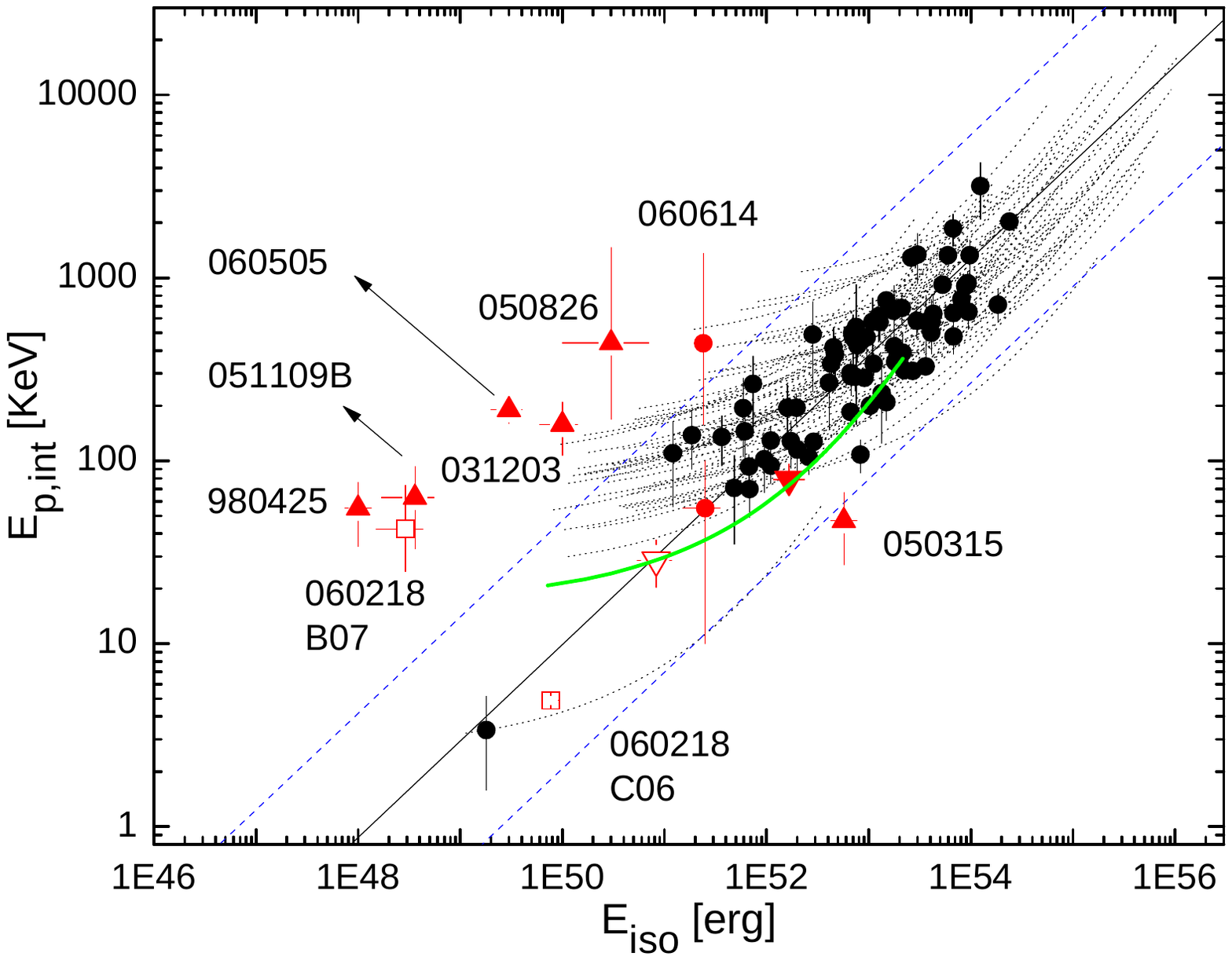}
\includegraphics[scale=0.31]{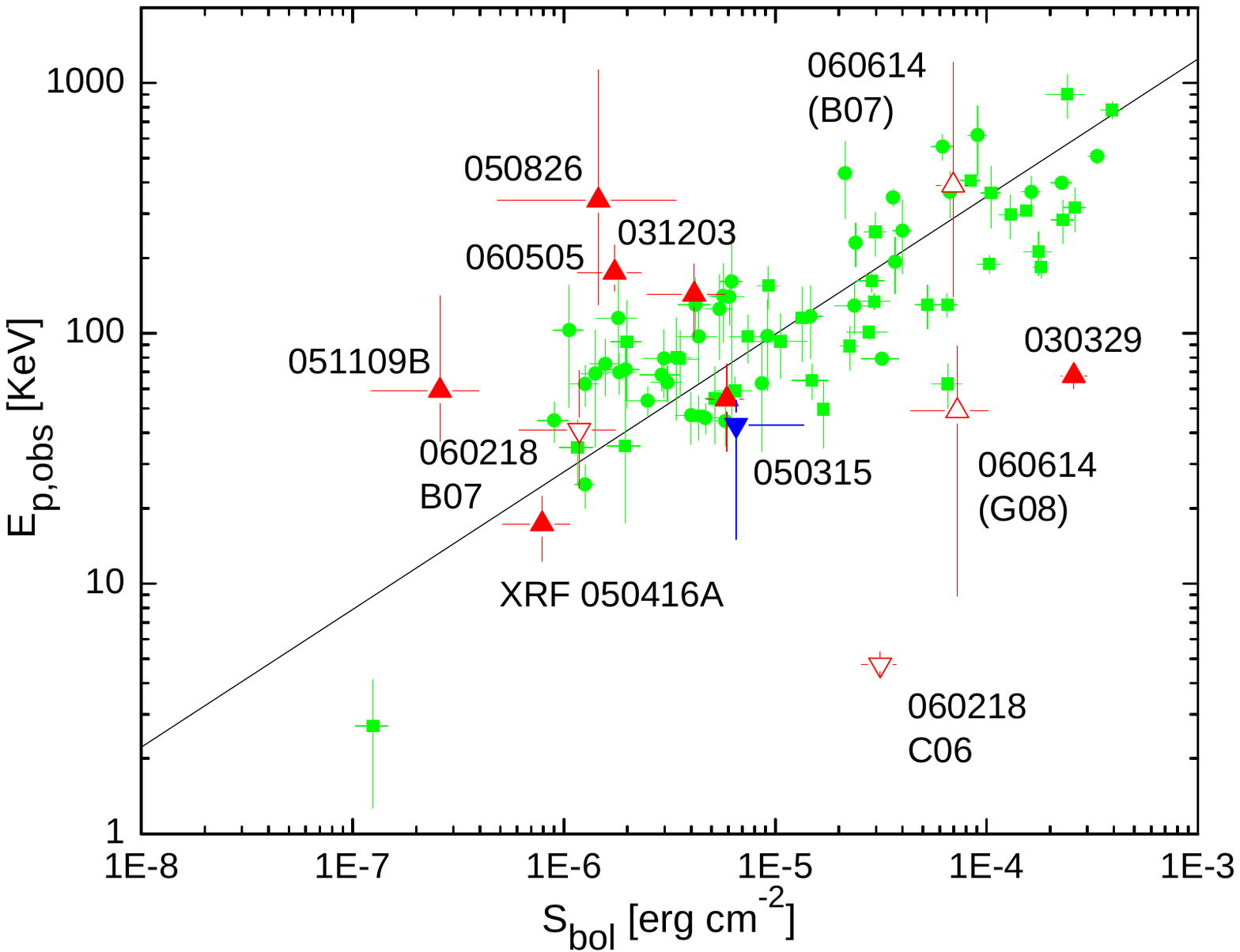}
\caption{ {\bf Left}: Plot of peak energy $\epi$ versus the isotropically emitted energy $\eiso$, with the G08 sample shown as black points.  The apparent correlation of these points is the basis for the Amati relation.  The dotted lines represent the trajectories of the LGRBs for different values of redshifts while holding all other GRB parameters fixed. The trajectories show the indifference of GRBs to redshift, indicating that {\bf the Amati relation is relatively insensitive to redshift}.  For any randomly given redshifts with $z>0.2$, all G08 LGRBs (black circles) reside within the $3\sigma$ consistency region of Amati relation -- the two blue dashed lines. In other words, {\bf there is no need to know the measured redshifts of LGRBs in G08 sample to define the tight Amati relation in the rest frame. One can almost always obtain a tight relation similar to Amati relation in the rest frame by attributing each burst a random redshift}. The head-down (empty red) triangle in the plot is XRF 050416A, claimed by Sakamoto et al. (2006) to be a further confirmation of the Amati relation in the gap of this relation around the 30-80 KeV range. The solid green line is the redshift trajectory of this burst holding all other parameters fixed. The consistency of this event with the rest frame Amati relation appears not to be due to a physical origin, but merely a result of lying along the same region in the observer frame where LGRBs define a strong observer frame correlation ({\it Left} plot) due to selection effects. For any value of redshift $z>0.09$, XRF 050416A is always consistent with the Amati relation at $<3\sigma$.  Interestingly, all Amati relation outliers except GRB 050315 \& 050826 have $z<0.2$. The head-down red-filled triangle is GRB 030329, consistent with the rest frame Amati relation, but an outlier to its observer frame counterpart. {\bf Right}: Plot of $\epo$ versus $\sbol$ for the G08 sample of LGRBs used to define the Amati relation in the rest frame plane. The green colored data points represent the same LGRBs (black circles) of the {\it Right} plot. The labeled bursts are LGRBs that are outliers to either the observer- or the rest- frame correlations as defined by the green and black circles in the corresponding plots, except XRF 050416A which is consistent with Amati relation in both planes. The head-up red triangle next to GRB 050315 is GRB 980425, a LGRB very well consistent with the observer plane Amati relation, but a far outlier to the rest frame relation. {\bf Both observer and rest frame Amati relations have comparable scatters (0.23 dex \& 0.21 dex respectively) and correlation coefficients that are not significantly different from each other ($\tau_{K,obs}=0.59\pm 0.05,7.5\sigma$ \& $\tau_{K,rest}=0.65\pm 0.04,8\sigma$ respectively)}. \label{AG08}}
\end{figure*}

The significant frequency, at over 19\%, of $>3\sigma$ outliers to the Amati relation found here is in contrast to the complete lack of outliers reported in G08 and the $\sim6\%$ outlier frequency reported in N08.  The presently reported outlier frequency is comparable to the $\sim 25\%$ outlier frequency reported by NP05a but still less than the $\sim 88\%$ outlier frequency reported by Band \& Preece (2005).  We find a similar outlier frequency to the $L_{iso}-\epi$ relation , $>8\%$ at $>3\sigma$, which is again in stark contrast to the findings of N08 with only $\sim 0.2\%$ at $>3\sigma$ outliers to this relation.  All of these discrepancies can be attributable to two sources: either an old version of the Amati relation was being used, or a different subsample of BATSE LGRBs was being used.  A third possibility raised initially by Ghirlanda et al. (2005a), that the apparent high frequency of outliers found by the method given in NP05a, could be due to the assumption of a low scatter in the Amati relation, is rejected. Were it true, the practical uses of the Amati relation, as well as other correlations intimately connected with, most importantly the Ghirlanda relation, would be limited. 

Even if the Amati relation as given by G08 is exact, it is not a significantly more accurate estimator of redshifts than random, since the positions of the bursts relative to these relations appear to be indifferent to a wide range of z. Figure~\ref{AG08} ({\it Left}) shows the \eiep ~of a sample of LGRBs used by G08 to define the Amati relation together with their trajectories on this plane for a wide range of redshifts ($0.2<z<20$).  Inspection of the plot indicates that the scatter and correlation strength of the relation depends very weakly on the redshifts of the bursts. The same sample of bursts (black dots in the {\it Left} plot of Figure~\ref{AG08}) also define a strong correlation in the observer plane (green dots in the {\it Right} plot of Figure~\ref{AG08}). The trajectories are however, different at lower redshifts and deviate from the Amati relation in \eiep ~plane for all bursts. This could well explain why almost all outliers to Amati relation have redshifts $z<0.2$. 

The Amati relation is created by apparently low dispersion, highly correlated bivariate distribution of GRBs in the \sbep ~plane.  The low dispersion effect is itself created by the detection and selection effects on the faint and low energy edges.  To show this more clearly, we derive the linear fits in the observer and rest frames to the sample of G08 bursts by excluding outliers in both planes as labeled in Figure~\ref{AG08} (except XRF 050416A which is not an outlier in either of the planes and was labeled for another reason to be discussed below), for which we find,

\begin{equation}
\label{eq:eqn2}
Log\left( \epo \right) = 4.75 + 0.55 Log\left( \sbol \right) , \\
\end{equation}
\begin{equation}
\label{eq:eqn3}
Log\left( \epi \right) = -25.11 + 0.52 Log\left( \eiso \right) . \\
\end{equation}

The variance between $\sbol$ and $\epi$ in both the observer and rest frames, as shown in the plots of Figure~\ref{AG08} has about the same scatter -- 0.23 dex and 0.21 dex -- with a slightly higher correlation coefficient being found in the rest frame of the bursts: $\tau_{K,obs}=0.59\pm 0.05, 7.5\sigma$ \& $\tau_{K,rest}=0.65\pm 0.04, 8\sigma$ respectively. This slight improvement, however, is statistically marginal, undermining a potential physical origin to the Amati relation. According to F-test, there is only $1\sigma$ ($p=0.285$) weak evidence of a significant difference between the variances of the observer and rest frame Amati relations.

In order to show how redshifting of the parameters on both sides of the relation Eq.~\eqref{eq:eqn2} can boost the existing correlation in the observer plane, we run a Monte Carlo simulation by giving each LGRB used to construct the relation ~\eqref{eq:eqn2} a random redshift taken from the sample over a large number of iterations taking into account of the limited energy budget of LGRBs, assuming $\Omega_{M}=0.27, \Omega_{\Lambda}=0.73$ \& $H_{0}=72$ Km$/$sMpc.  Averaging over all iterations we find that the Kendall's rank correlation coefficient of the observed sample is generally enhanced by $\sim 0.05$, that is about the same as the correlation coefficients difference ($\sim 0.06$) obtained for the sample of bursts with real measured redshifts. Also, 53\% of the iterations result in a median scatter in the rest frame distribution that is smaller than the median scatter of the observer frame.

\begin{figure}
\includegraphics[scale=0.31]{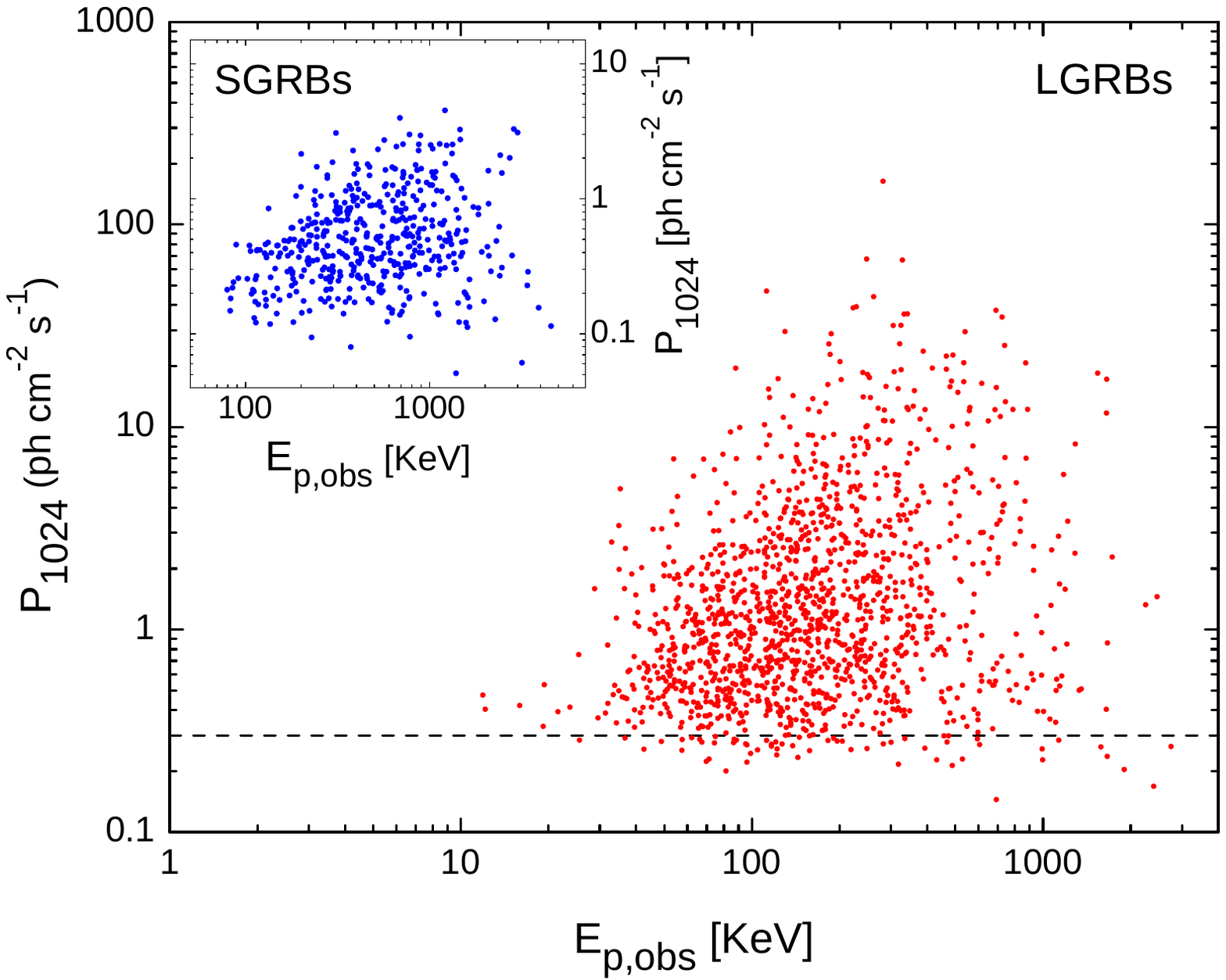}
\caption{A plot of 1-second peak flux $P_{1024}$ versus $\epo$ for 1900 BATSE GRBs.  The main plot contains only LGRBs classified via a fuzzy clustering algorithm described in Shahmoradi (2010), while the inset plot contains only SGRBs.  The dashed line represents the nominal BATSE trigger threshold on $1024ms$ timescale. The lack of very soft bright bursts is visible in the top left of both plots which seems to have a physical origin, while there are apparently no limits for the positions of the bursts on the dim hard side of the plots except BATSE trigger threshold. \label{pbcntepk}}
\end{figure}

Inspection of Figures~\ref{outliers} ~\& \ref{AG08} ({\it Left} plot), however, indicates a lack of bursts on the lower right sides of the observer \& rest frame planes of \sbep ~\& \eiep.  In other words, GRBs that are both bright and soft appear to be rare. This could have a physical origin and would likely be unaffected by detection threshold limits.  In our estimation, {\bf the existence of a probable physically-based upper limit for the hardness ($\epo$) of GRBs as a function of fluence, together with the selection effects, have possibly led to the creation of the low-dispersion Amati relation, and similar $\epi$-based correlations such as the Ghirlanda relation, in the rest frame of the bursts} (Figure~\ref{pbcntepk}).

It is interesting to wonder `{\it why almost all of the outliers to the Amati relation appear to be sub-energetic, lying at redshifts of $z<0.2$?}'  Inspection of Figure~\ref{AG08} ({\it Left}) indicates a likely reason: $\epi$ is insensitive to redshift when the redshift is small. Specifically, any burst that lies within $3\sigma$ limits of this relation in the observer plane with a redshift $z>0.2$, will always be consistent with the Amati relation (Eq.~\eqref{eq:eqn3}) in the rest frame at $<3\sigma$ regardless of the any possible redshift that the burst might have. On the other hand, any burst consistent with relation~\eqref{eq:eqn2} at $<3\sigma$ with a redshift $z<0.2$ would almost always be a certain outlier to Amati relation at $>3\sigma$ in the rest frame, or in reverse, any burst with $z<0.2$ that is consistent with Amati relation, would almost always be a certain outlier to relation~\eqref{eq:eqn2} at $>3\sigma$. Therefore, the apparent deviation from the Amati relation observed for sub-luminous bursts is unphysical and created by the random redshifts of the bursts. It is also of no surprise that the newly detected bursts are generally consistent with this relation, so long as they are detected and selected in the region of consistency with the observer frame Amati relation with a $z>0.2$.

As an example, we consider XRF 050416A that is reported by Sakamoto et al. (2006) to be a further ``confirmation" of the inclusiveness of the Amati relation in the gap around the 30-80 KeV range.  The redshift trajectory of this burst is shown in Figure~\ref{AG08} ({\it Left} plot), holding all other parameters fixed (the green solid line in the graph). Given any random redshifts ($z>0.1$), this burst is always consistent with Amati relation at $<3\sigma$, indicating a nonphysical origin for its consistency.

It is interesting that there are no nearby ($z<0.2$) bright outliers to Amati relation which is possibly due to the luminosity distribution and the redshift evolution of luminosity function of the bursts (e.g. Kocevsky \& Liang 2006; Lloyd-Ronning et al. 2002) as well as possible yet unknown effects of jet opening angles of the bursts.  In other words, {\bf the lack of bright bursts detected in the very nearby universe results in the apparent tightness of the Amati relation at high isotropic energies ($\eiso \gtrsim10^{52}$[ergs]) as well as a large scatter with frequent outliers at lower isotropic energies due to a general insensitivity of $\epi$ to redshift at $z<0.2$}.
   
\section{Concluding Remarks}
\label{sec:conclusion}
Throughout the presented analyses, we investigated the effects of BATSE Large Area Detectors' threshold limitations on triggering GRBs over a wide range of spectral peak energies ($\epo$). In order to compare the distribution of BATSE GRBs with these triggering thresholds on the plane of bolometric fluence vs. peak energy, we relied on the $hardness-\epo$ correlation reported by Shahmoradi \& Nemiroff (2010{\it a}) to map the hardness vs. bolometric fluence (\sbhr) plane of 1900 BATSE GRBs -- for which accurate continuous light curve data as well as fluence in all energy channels were available -- into the plane of  \sbep. 

The results of the simulation indicate that the distribution of GRBs on \sbep ~plane is affected by the BATSE trigger thresholds on the regions where dim hard bursts reside. This is simply due to the fact that BATSE specifically and GRB detectors generally are photon counters rather than bolometers. Therefore, for a given bolometric peak flux (e.g. in units of $ergs~cm^{-2}~s^{-1}$), the harder bursts would have less photon count rates than the softer GRBs and consequently, less chance of detection. Although the significance of the BATSE triggering thresholds on the bivariate distribution of LGRBs in \sbep ~plane has yet to be determined (Shahmoradi \& Nemiroff 2010{\it b}), the current sample of LGRBs detected by BATSE provides evidence against the proposed low-dispersion Amati \& $L_{iso}-\epi$ relations (Yonetoku et al. 2004; Amati 2006):

About 20\% of the BATSE sample of LGRBs considered here appear inconsistent with the Amati relation at $>3\sigma$ based on the traditional ($\tn >3 ~[sec]$) or fuzzy clustering classification (Shahmoradi 2010) of BATSE LGRBs (\S \ref{sec:sbep}). The fraction of outliers is likely more than the ratio obtained for reasons of sample inclusion and statistical methodology.  The current sample would likely include more outliers were some GRBs themselves not excluded by detection thresholds of BATSE LADs. In addition, the method of NP05a used herein can only set a lower limit to the number of the outliers. This is bolstered, knowing that none of the currently known outliers to the Amati relation with firmly measured redshifts, except GRB 050826, could be flagged as certain outliers at $>3\sigma$ by this method. 

Similar conclusions also hold for the simulations of BATSE trigger thresholds on the plane of \pbep ~(\S \ref{sec:pbep}). Using the same method as applied for \sbep ~plot of BATSE GRBs, we can set a lower limit on the inconsistency of BATSE LGRBs with $L_{iso}-\epi$ relation. We find that at least 8\% of the bright BATSE LGBRs are certain outliers to this relation at $>3\sigma$ as given recently by G09. 

It is notable that the results of our simulations are in contrast with the findings of previous authors, in particular N08 \& G08, where they find that BATSE trigger threshold, as well as spectral analysis limits on the two planes of \sbep ~\& \pbep ~have possibly little or no effects on the distributions of BATSE LGRBs on these two planes. The reason for the discrepancies should be sought in the values of the limiting parameters that they use in their simulations, such the average ratio of fluence to peak flux of the bursts (FPR) and the nominal durations of the bursts considered therein, all taken from a small fraction of {\it detected}, {\it spectrally analyzed} BATSE LGRBs. In this sense, their analyses suffer from a circular logic problem (\S \ref{sec:sbep}). In addition, the use spectral models with fixed photon indices in their simulations, results in a severe underestimation of the selection effects (\S \ref{sec:pbep}). The strong evolution of the peak-energy ($\epo$) in the light curves of the bursts (e.g. K06, Ryde 1999; Crider et al. 1999; Band 1997; Crider et al. 1997; Liang \& Kargatis 1996; Ford et al. 1995) is another factor that is overlooked in the simulations of G08 \& N08.  We have worked to make the simulations presented here free from the above mentioned deficiencies.

Another strong argument that favors an unphysical origin for the Amati, Ghirlanda and possibly other 3-parameter relations, such as the empirical Liang-Zhang relation, comes from inter-comparisons between the proposed relations.  Previous authors have reported a significant scatter reduction in transforming the Amati relation to these 3-parameter relations, specifically the Ghirlanda relation. However, considering the same sample for both relations that are being compared to each other, we find that the scatter reduction and correlation improvements are insignificant (\S \ref{sec:Ghirlanda}). Therefore, in order to have a meaningful comparison of any two relations with each other, in particular the Amati \& Ghirlanda relations, it is important to consider the same data set for both relations.

It is also noteworthy that the sample of LGRBs considered by G08 to construct the tight Amati relation also shows a strong correlation in the observer frame, with a scatter comparable to the dispersion in the rest frame Amati relation, differing by only 0.02 dex. This indicates that the tightness of the Amati relation is only a ghost of the tight correlation of this sample of LGRBs in the observer frame, reinforced by redshifting of the spectral parameters of the bursts from the observer to the rest frame plane. For any random redshifts that these bursts might have, the rest frame Amati relation is on average always tighter than the Amati relation in the observer frame (\S \ref{sec:discussion}). Moreover, the apparent frequent inconsistencies of the sub-luminous LGRBs with the Amati relation, appear to have no physical origin and can be attributed purely to redshifting of the spectral parameters of the bursts that mainly reside on a narrow strip in the observer frame, by a redshift $z<0.2$ (\S \ref{sec:discussion} \& Figure~\ref{outliers},~\ref{AG08}).

In sum, the Amati relation as proposed by Amati (2002), Amati (2006) \& Ghirlanda et al. (2008) appears to be greatly affected by complex selection effects in triggering, spectral analyses \& redshift measurements of LGRBs on the dim side of the \sbep ~plane.  The lack of LGRBs on the soft bright side of the \sbep ~plane might possibly retain an underlying physical origin.  Nevertheless, the practical use of $\epo$ as a standard candle is questioned, as its detector convolutions likely compromise its use as a discerning probe for cosmological models.
\\
\\
\\
This work could not have been accomplished without the vast time and efforts spent by many workers over the past decade, in particular BATSE team, including the designers, builders, and analysts for the burst detectors on board the Gamma Ray Observatory who have accumulated and analyzed the observations and summarized them in BATSE GRB catalogs.   In particular, we acknowledge several useful communications with David Band, and dedicate this paper to his memory.

\label{lastpage}

\end{document}